\documentclass[lettersize,journal]{IEEEtran}
\usepackage{amsmath,amsfonts}
\usepackage{amsthm,amssymb}
\usepackage{mathrsfs}
\usepackage[ruled,linesnumbered]{algorithm2e}
\usepackage{array}
\usepackage[caption=false,font=normalsize,labelfont=sf,textfont=sf]{subfig}
\usepackage{textcomp}
\usepackage{stfloats}
\usepackage{url}
\usepackage{verbatim}
\usepackage{graphicx}
\usepackage{cite}
\usepackage{array}
\usepackage{booktabs}
\usepackage{makecell} % 表格内容换行
\usepackage{multirow} % 表格多行
\usepackage{colortbl}
\usepackage{color}
\usepackage{threeparttable} % 表格下方注释
\usepackage{diagbox} % table cell in theagonal
\usepackage{orcidlink} % orcid链接
\hypersetup{hidelinks} % hide the hyperlink box
\begin{document}

\title{PromptSAM+: Malware Detection based on \\ Prompt Segment Anything Model}

\author{Xingyuan Wei \orcidlink{0009-0001-6595-4222},
Yichen Liu,
Ce Li,
Ning Li,
Degang Sun,
Yan Wang$^{\ast}$ 
        % <-this % stops a space
%\IEEEauthorrefmark{}
\thanks{* Corresponding author}% <-this % stops a space
}
% \thanks{\& These authors contributed equally to this work and should be considered co-first authors.}}

\markboth{Journal of \LaTeX\ Class Files,~Vol.~14, No.~8, August~2021}
{Shell \MakeLowercase{\textit{et al.}}: A Sample Article Using IEEEtran.cls for IEEE Journals}


\maketitle
%  ———————————————————----———————Abstract———————————————
\begin{abstract}
% 机器学习和深度学习(ML/DL)已经被广泛应用于检测恶意软件，现有的一些检测方法性能十分的强大。但是，目前恶意软件检测领域仍然存在几个问题:(1)现有的工作过度在意准确度而忽略了实用性，很少有工作把误报率和漏报率作为一个重要指标(2)考虑到恶意软件的演变，分类器的性能会随着时间的推移而显著下降，恶意软件检测器实用性大大降低。(3)先前基于ML/DL的工作都过度依赖于足够的标记数据来进行模型训练，这些技术大多依赖于特征工程工作或领域知识来构建特征数据库,如果正确的标签稀缺,现有的恶意软件中可能会很脆弱。随着计算机视觉领域的发展，基于视觉的恶意软件检测技术也在迅速发展。在本文中，我们提出了一个可视化的恶意软件通用增强分类框架`PromptSAM+',它是基于一种\textbf{Prompt} \textbf{S}egment \textbf{A}nything \textbf{M}odel的大视觉网络分割模型，我们的实验结果表明，‘PromptSAM+’在恶意软件检测和分类方面是有效和高效的，具有较高的分类准确率以及比较低的误报率和漏报率，所提出的方法在几个数据集上优于最先进基于图像的恶意软件检测技术。‘PromptSAM+’可以的对现有的基于恶意软件图像分类器老化的减缓，可以减轻标记新的恶意软件样本时节省了主动学习所需的大量人力。我们进行了Windows和Android两个平台数据集实验，均取得了好的效果。同时我们完成消融实验，在几个数据集上的结果表明我们的模型找到了大视觉网络起作用的模块。
Machine learning and deep learning (ML/DL) have been extensively applied in malware detection, and some existing methods demonstrate robust performance. However, several issues persist in the field of malware detection: (1) Existing work often overemphasizes accuracy at the expense of practicality, rarely considering false positive and false negative rates as important metrics. (2) Considering the evolution of malware, the performance of classifiers significantly declines over time, greatly reducing the practicality of malware detectors. (3) Prior ML/DL-based efforts heavily rely on ample labeled data for model training, largely dependent on feature engineering or domain knowledge to build feature databases, making them vulnerable if correct labels are scarce. With the development of computer vision, vision-based malware detection technology has also rapidly evolved. In this paper, we propose a visual malware general enhancement classification framework, `PromptSAM+', based on a large visual network segmentation model, the \textbf{Prompt} \textbf{Segment} \textbf{Anything} \textbf{Model}(named PromptSAM+). Our experimental results indicate that 'PromptSAM+' is effective and efficient in malware detection and classification, achieving high accuracy and low rates of false positives and negatives. The proposed method outperforms the most advanced image-based malware detection technologies on several datasets. 'PromptSAM+' can mitigate aging in existing image-based malware classifiers, reducing the considerable manpower needed for labeling new malware samples through active learning. We conducted experiments on datasets for both Windows and Android platforms, achieving favorable outcomes. Additionally, our ablation experiments on several datasets demonstrate that our model identifies effective modules within the large visual network.
\end{abstract}


% —————————————————---——————Introduction———————————第一段————————————————
% 恶意软件（malicious software）攻击日益增多，已经成为全球互联网用户面临的一项严峻挑战。目前，恶意软件作为网络犯罪分子执行恶意活动的主要手段之一。根据AV-Test统计报告（AV-Test恶意软件统计，2024年），大约有1亿新的恶意软件文件在2024年上半年被发现\cite{AVTestMalware}。这些包括各种类型的恶意软件，如PowerShell威胁、MacOS恶意软件、Office恶意软件、移动Android恶意软件和Linux恶意软件。此外，每年新恶意软件家族及其变种的出现根据前述报告造成了惊人的损失。根据最新报告，网络犯罪将在2024年每年给世界造成9.5万亿美元的损失\cite{einpresswireReportCybercrime}。这些统计数据证实了恶意软件对互联网用户构成不断增长的威胁。因此，在恶意软件各种变种和重大危险的背景下，寻找一种跨平台、通用的检测方法变得至关重要。
\section{Introduction}
\IEEEPARstart{T}{he} ever-increasing threat of malware (\textbf{mal}icious soft\textbf{ware}) attacks has emerged as a formidable challenge for internet users worldwide. Presently, malware stands as a principal vector employed by cybercriminals to execute malicious activities. According to the AV-Test statistics report (AV-Test malware statistics, 2024), approximately 100 million new malware files were discovered in the first half of 2024. These included various types of malware such as PowerShell threats, MacOS malware, Office malware, Mobile Android malware, and Linux malware \cite{AVTestMalware}. Besides, the emergence of new malicious families and their malware variants every year causes astonishing damage according to the aforementioned reports. According to the latest report, cybercrime is to cost the world \$9.5 trillion USD annually in 2024 \cite{einpresswireReportCybercrime}. These statistics support the fact that malware is a growing threat to Internet users. Therefore, in the context of malware's numerous variants and significant dangers, finding a cross-platform, universal detection method becomes crucially important.

%—————————————————-----—————————第二段—————————————————————————
% 越来越多的研究表明，将机器学习和深度深度学习的方法应用到恶意软件检测领域是一种有效且有前途的，机器学习和深度学习已经通过让模型从大量数据中学习，彻底改变了网络安全领域。现有的文献表明，使用基于ML/DL技术的恶意软件分类器通常可以达到相当高的性能，检测准确率高达99%，几乎没有留下进一步改进的空间\cite{razgallah2021survey} \cite{liu2022deep} \cite{qiu2020survey}。然而，这些高性能方法在实践中似乎不太可行，因为恶意软件防御仍然是一个具有挑战性的问题，恶意应用程序继续对人们构成越来越大的威胁。恶意软件检测器在大规模筛查恶意软件的过程中，不能只专注于其准确率而忽略了漏报率和误报率，这一点至关重要。如果漏报率过高，那么恶意软件可能会未受阻碍地运行，对系统造成严重危害；而如果误报率过高，则可能导致大量良性软件被错误地阻断，影响用户体验并减少系统的可用性。因此，在开发和评估恶意软件检测系统时，不仅要考虑多个性能指标，也要考虑一定的实用价值，以确保系统在正确识别恶意软件的同时，也能最大程度地减少对正常操作的影响，这通常需要通过优化检测算法和使用更加精细化的特征来实现平衡。
Increasing research indicates that applying machine learning and deep learning approaches to malware detection is effective and promising. Machine learning and deep learning have revolutionized the field of cybersecurity by enabling models to learn from vast amounts of data. We find that ML/DL techniques can often achieve remarkably high performance, with detection accuracies reaching up to 99\%, leaving little room for further improvement \cite{razgallah2021survey} \cite{liu2022deep} \cite{qiu2020survey}. However, these high-performance methods appear less feasible in practice because malware defense remains a challenging issue, and malicious applications continue to pose an increasing threat to people. There is an excessive focus on the accuracy of malware detection, which often overlooks its practicality. Malware detectors in the process of screening malware on a large scale, must not solely focus on accuracy while neglecting false negative and false positive rates, which are crucial. A high false negative rate could allow malware to operate unimpeded, causing severe damage to systems; conversely, a high false positive rate might lead to a large number of benign applications being wrongly blocked, affecting user experience and reducing system usability. Therefore, when developing and evaluating malware detection systems, it is essential not only to consider multiple performance metrics but also to take into account the practical utility to ensure that the system can correctly identify malware while minimizing the impact on normal operations. This typically requires balancing through the optimization of detection algorithms and the use of more refined features.

% ———————————————————---——--—------———————第三段————————————----————————
% 恶意软件检测领域中一个比较重要的问题是恶意软件的进化，即恶意软件检测器检测能力退化。利用ML/DL算法提取特征构建一个恶意软件分类器，这些恶意软件检测器在刚开始的时候性能表现卓越，可以达到良好的分类效果，但是随着时间的推移性能变得极差，这类问题类在有些文献称为时间衰减\cite{pendlebury2019tesseract}, 模型降解\cite{lei2019evedroid}和退化\cite{droidSpan}。MaMaDroid\cite{mamadroid2016}提出了一种能够适应Android规范变化的检测方法。具体来说，MaMaDroid首先将应用程序编程接口(API)抽象为每个Android应用程序派生的API执行路径中的相应包(或包族)。然后将所有抽象路径总结为马尔可夫模型，并将马尔可夫模型转换为模型训练和测试中每个应用程序的特征向量。然而，API之间的关系比包之间的关系要广泛得多。APIGraph\cite{zhang2020APIGraph}的框架通过software development kits软件开发工具包(SDK)提供的文档构建API之间的联系，从关系图中提取到语义，通过语义相似性增加抗漂移能力。诚然mamadroid,APIGraph等这些方法确实能解决掉一些问题，但是很难直接在现有的方法上进行扩展，在不同平台需要单独设计出一套相应的解决方案，现有的方案很难具有通用性和复用性。因此，我们本文的工作试图提出一套可以直接附着在现有恶意软件图像分类器的增强方法，能够弥补当前问题的一个通用框架。
An important issue in the field of malware detection is the evolution of malware, namely the degradation of the detection capabilities of malware detectors. Using ML/DL algorithms to extract features and construct a malware classifier, these malware detectors initially perform excellently, achieving good classification results. However, over time their performance deteriorates significantly, a problem referred to in some literature as temporal decay \cite{pendlebury2019tesseract}, model degradation \cite{lei2019evedroid}, and degradation \cite{droidSpan}. MaMaDroid \cite{mamadroid2016} proposes a detection method that adapts to changes in Android specifications. Specifically, MaMaDroid first abstracts the Application Programming Interface (API) into the corresponding package (or package family) in the API execution paths derived from each Android application. It then summarizes all abstracted paths into a Markov model and converts the Markov model into feature vectors for each application during model training and testing. However, the relationships among APIs are much more extensive than those among packages. The APIGraph \cite{zhang2020APIGraph} framework constructs connections among APIs through documentation provided by software development kits (SDKs), extracts semantics from the relationship graph, and enhances drift resistance through semantic similarity. Indeed, methods like MaMaDroid and APIGraph can solve some issues, but they are difficult to directly extend on existing methods, requiring separate tailored solutions for different platforms, making existing solutions lack generality and reusability. Therefore, our work in this paper attempts to propose a universal framework that can be directly attached to existing malware image classifiers as an enhancement method, addressing current issues.

With the advancement of computer vision, image-based malware detection technology is rapidly evolving. This visualization method intuitively displays the characteristics of different types of malware, providing a new pathway for malware analysis. The earliest instance of image-based malware detection dates back to 2011, proposed by Nataraj et al.\cite{first_malware_image}, who represented malware bytes as pixels in an image. They converted the binary information of malware into images and then extracted GIST features for classification, ultimately utilizing k-nearest neighbors (kNN) classification technology on the Malimg\cite{first_malware_image} dataset. Malware detection using deep learning visual models such as VGG16 \cite{vggForMalware2018}, CNN+LSTM \cite{cnnLstmForMalware2019}, DenseNet\cite{DensenetForMalware2021}, Xception\cite{xceptionForMalare2019}, ResNet\cite{resnetForMalware2021}, \cite{resnextForMalware2020}, AlexNet\cite{alexnet2016} has significantly enhanced the speed and performance of malware discovery and detection. However, current vision-based methods still face several challenges, such as the rudimentary nature of existing vision-based detection methods that too bluntly transfer existing visual models directly to malware detection without adaptively adjusting for the task. Issues including how to convert malware into images without losing information and how to capture useful malware features through the perceptual field of image channels to further improve malware detection and classification performance also need to be researched.

%—————————————————————————第六段——————————我们是怎么解决这个问题的——————————
% 为了解决上述的问题，在本文的研究中我们把大视觉网络Segment Anything Model(SAM)模型的语义纳入进来，获得了视觉模型的语义信息。我们找到了大视觉网络真正起作用的部分，我们开创性地利用了大视觉模型的语义信息提升恶意软件的检测和分类任务的表现。我们的模型可以找到恶意软件家族的共同特性，减小恶意软件检测的误报率和漏报率，增强了实用性，从而减少分析工作量，加快恶意软件的分析速度。
To address the issues mentioned above, our research incorporates the semantics of the large visual network, the Segment Anything Model (SAM), to harness the semantic information of visual models. We identified the truly effective components of the large visual network and innovatively utilized the semantic information from the large visual model to enhance the performance of malware detection and classification tasks. Our model can identify common characteristics among malware families, reduce the false positive and false negative rates in malware detection, and enhance practicality, thereby reducing the analytical workload and accelerating the analysis of malware.

%此外，我们还证明了大视觉模型作为一种语义分割图像模型所包含的语义信息在恶意软件检测方面具有一定的可解释性。
% 我们首先评估了我们的'PromptSAM+'在大规模数据集上的表现。对于恶意软件家族，恶意软件家族地辨认可以发现同一家族的相似性，找到通用的恶意代码片段，因此我们根据大视觉网络的特性进行改进适用于恶意家族分类任务，我们的‘PromptSAM+’在家族数据上训练的heatmap也提示该模型能够识别出恶意软件家族的共性。我们评估了模型在多个benchmark数据集中家族分类的能力。我们将'PromptSAM+'与其他的视觉模型进行比较，例如传统经典的CNN, ResNet18, Resnet50\cite{he2016deep}等, 以及目前在基于图像检测恶意软件领域中最stata of the art(SOTA)的ResNeXt+ \cite{resNeXt+2024}，通过比较性能表现判断PromptSAM+是否更加有效。我们还探究了PromptSAM+的跨平台处理能力，探究分别在Windows和Android数据上表现情况。我们利用大视觉网络在隐藏的网络语义信息，确实可以对现有的视觉模型通用性的增强。（这段需要改 补充的我们的亮点）
%******************************废料段*************************************************************
% 自动化工具和方法可能会重用一些模块来开发恶意软件变体，因此这些重用的模块可以用于对恶意软件进行分类或识别恶意软件家族。因此，可以分析恶意软件变体之间可能存在的相似性\cite{han2015entropyGraphs} 这段不翻译%
%1. 现阶段生成图像的存在的问题，当前生成图像的方式
%2. 图像生成方法的问题，大问题是所有的方法都抗漂移不好，抗漂移模型的通用性并不好
%这一段应该讲本文的独创性：用了视觉模型的语义信息，并且给出了一定的可解释性；造了2个数据集。由于目前软件对抗和软件专利保护经常使用混淆等技术，因此混淆技术导致很多软件被检测系统漏报或误报，我们也验证我们方法面对对混淆技术时候的Robust。我们利用分割网络SAM的语义信息用于malware图像的分割检测，我们的heatmap图像证明了对于同一类恶意软件我们的模型确实抓住了语义相同的部分。
%近年来，还出现了一些基于自然语言处理技术(NLP)和图神经网络(GNN)的方法进行恶意软件检测，也取得不错的效果\cite{zhang2020APIGraph}, \cite{bai2023argusdroid},\cite{he2022msdroid}。
%******************************废料段*************************************************************

% ————————————————————----—————————第五段：贡献—————————————————————————
\textbf{Contributions.} The main contributions of our work are summarized as follows:
\begin{enumerate}
% 我们提出一个通用性的框架名为PromptSAM+，我们我们的模型具有良好的泛化能力，可以很好的适用于恶意软件检测任务，并且实现多平台检测，既可以检测Android恶意软件也可以检测Windows恶意软件。我们提出的模型它能够很好的区分恶意软件和良性软件，我们在155k个数据集中做了充分的实验，验证我们提出的模型的有效性和实用性。同时，我们的模型是一个通用的检测框架，这个框架可以直接增强现有的视觉网络，可以当成一个模块很方便快速直接插入到现有视觉模型中，整体提高模型性能。

\item{We propose a versatile framework named PromptSAM+, which demonstrates strong generalization capabilities suitable for malware detection tasks across multiple platforms, capable of detecting both Android and Windows malware. Our model effectively distinguishes between malicious and benign software. We conducted extensive experiments with a dataset of 155k samples to validate the efficacy and practicality of our proposed model. our method has shown robust detection and classification capabilities. Additionally, our model serves as a general detection framework that can directly enhance existing visual networks. It can be conveniently and rapidly integrated as a module into existing visual models, significantly improving overall model performance.}

% 首次将大视觉模型中的语义信息利用到恶意软件检测中。在具体实验中,我们将SAM分割模型的语义信息融入到我们的框架里。我们是第一个使用SAM模型在恶意软件方面上。我们评估模型了模型的检测能力，恶意软件家族分类能力，Few-shot能力（its few-shot learning capacity 这个没有篇幅写了），抗概念漂移能力，模型在保持较高性能的同时也能显著地（significantly reduces）延缓图像恶意软件模型分类器的衰退。
\item{We are the first to utilize semantic information from large visual models in malware detection. In our experiments, we incorporated semantic information from the SAM segmentation model into our framework. This is the inaugural use of the SAM model in the context of malware. We evaluated the model's detection capabilities, its ability to classify malware families, and its resistance to concept drift. While maintaining high performance, the model also significantly reduces the degradation of the classifier in image-based malware models.}

% 我们修改了以往Android恶意软件图像的生成方法，提出了一种新的Android恶意软件图像生成方法，通过剪切对齐融合的手段把Android恶意软件转换成图像，利用图像模型的快速处理能力，使得我们的方法可以对实现世界中的恶意软件进行大规模的通用筛查过滤。

\item{We have modified the traditional method of generating images for Android malware and proposed a new method for creating Android malware images. By employing a technique of cutting, aligning, and fusing, we convert Android malware into images. This leverages the rapid processing capabilities of image models, enabling our method to perform large-scale, general screening and filtering of malware in the real world.}

% 我们利用VirusTotal\cite{virustotal}Euphony\cite{Euphony}等工具，基于Androzoo\cite{androzoo}, MalNet\cite{malnet}几个开源数据库，收集了包含24个家族，数量达到27,294个的恶意软件家族数据集Prompt-Family, 以及2015年至2021年跨时间长度为7年数量为105,000的恶意软件数据集Prompt-Time。由于版权问题和防止恶意软件被错误利用等原因，我们不公开源文件数据，但是我们公开所有的恶意软件的图像数据作为科学研究使用。
\item{We utilized tools such as VirusTotal \cite{virustotal} and Euphony \cite{Euphony}, and based on open-source databases like Androzoo \cite{androzoo} and MalNet \cite{malnet}, we have collected a malware family dataset, Prompt-Family, comprising 27,294 samples across 24 families, and a malware dataset, Prompt-Time, which spans seven years from 2015 to 2021 and includes 105,000 samples. Due to copyright issues and to prevent the misuse of malware, we do not publicly release the source data files. However, we make all malware image data available for scientific research purposes\footnote{https://github.com/XingYuanWei/PromptSAM}.}
\end{enumerate}

% ———————————————————————————————————————第六段：架构—————————————————————————————————————————————————
% 这里讲述一下文章的架构
% The rest of the paper is organized as follows. Section \ref{related_work} discusses the related works on various malware classification techniques. Section \ref{methodology} provides the proposed framework and detailed proposed methods. Section \ref{expriments_and_result} describes some details of the experimental setup and analyzes the details of the results. Section \ref{discussion} provides discussion details of our method. Finally, Section \ref{conclusion_and_future} presents the conclusions and future work.
The rest of the paper is organized as follows. Section \ref{related_work} discusses the related works on various malware classification techniques. Section \ref{methodology} provides the proposed framework and detailed proposed methods. Section \ref{expriments_and_result} describes some details of the experimental setup and analyzes the details of the results. Finally, Section \ref{conclusion_and_future} presents the conclusions and future work.

\section{RELATED WORk}\label{related_work}
% In this section, we provide an overview of popular existing malware malware classification methods,
% Existing image-based malware detection and classification approaches are primarily based on conventional machine learning or deep learning with Convolutional Neural Networks.

\subsection{Problem Statement}
% 提出恶意软件检测的动机是恶意软件数据有限，为了利用这些有限的数据进行有效地检测，并且可以对多个平台的恶意软件进行检测，所以训练一个分类模型。因此，在这项工作中，我们解决了构建基于恶意软件的有效模型的问题。我们将恶意软件的检测和分类来定义如下:
\textit{Malware Detection And Family Classification:} The motivation for proposing a malware detection method arises from the limited availability of malware data. To effectively utilize this scarce data for detection, and to enable the identification of malware across multiple platforms, we have trained a classification model. Consequently, in this work, we have addressed the challenge of developing an effective model based on malware. The problem is constructed as follows:

% \begin{equation}
% \label{dataset_definition1}
% \mathcal{D} = \{{B + M + M_{ST}}\}
% \end{equation}

\begin{equation}
\label{dataset_definition2}
\mathcal{D} = \{{B + M + B_O + M_O + M_{ST}}\}
\end{equation}

Given an Android or Windows application, a malware detection system should be able to distinguish whether it is malicious or benign \cite{he2022msdroid}. Assume there is a dataset $\mathcal{D}$ containing $n$ samples (\textit{$D_1$, $D_2$, $D_3$.., $D_n$}). The distribution of real-world data is complex. In addition to benign software $B$ and malicious software $M$, there are also randomly mixed malicious and benign software, $M_O$ and $B_O$, as shown in equation \eqref{dataset_definition2}. The dataset $\mathcal{D}$ has a sufficient number of samples $n$ to train an effective malware detection model. However, only a few samples of complex data are available for training. Despite the limited complex data available for training machine learning models, a technique still needs to be designed to effectively differentiate these limited observed malicious and benign data.

\begin{equation}
\label{detection_def}
\textbf{\textit{C:}}(D_l) \Rightarrow \{{B,M}\}
\end{equation}

% 正如公式3中定义的那样，给定一个数据集D_l,
% 在恶意软件家族的分类任务中，数据集只包含恶意软件，没有其他良性软件，正如公式（4）中所示，其中M_F1,M_F2...M_Fn表示的恶意软件家族1,家族2至家族n，恶意软件家族分类的目的是在训练集D_l上训练一个分类器, 把各个家族的恶意软件区分开，正如公式(5)所示.
As shown in equation \eqref{detection_def}, given a dataset $D_l$, $D_l\subseteq \mathcal{D}$ containing a limited number of complex malware and benign samples, a classifier \textbf{\textit{C:}}must be built which can efficiently classify between limitedly seen classes, such as benign \textbf{$B$}, malware \textbf{$M$}. In the task of malware family classification, the dataset contains only malicious software and no benign software, as shown in Equation \eqref{family_classification_dataset}. $M_{F1}$, $M_{F2}$, $...$, $M_{Fn}$ represent malware families 1, 2 through to n, respectively. The objective of malware family classification is to train a classifier on the training dataset $D_l$, $D_l\subseteq \mathcal{D}$ which can distinguish between the malicious software of each family, as demonstrated in equation \eqref{family_classification_def}.
\begin{equation}
\label{family_classification_dataset}
\mathcal{D} = \{{M_{F1} + M_{F2} + M_{F3}+...+ M_{Fn}}\}
\end{equation}

\begin{equation}
\label{family_classification_def}
\textbf{\textit{C:}}(D_l) \Rightarrow \{{M_{F1}, M_{F2}, M_{F3},...,M_{Fn}}\}
\end{equation}

\textit{Concept Drift and Model Aging:}
% 概念漂移是机器学习中的一种常见现象，其中样本的统计属性随着时间而变化。基于机器学习的方法来说，所谓概念漂移指的是测试样本的统计特性随着时间的推移以不可预见的方式变化的现象，一般情况下随着时间的推移，模型的检测精度将降低。与猫狗分类这些传统基于特征分类任务不同是，猫和狗的特征十分的稳定，几乎不会随着时间变化而变化。由于Android和Windows等操作系统为了大量提高市场的占有率，系统版本的高频率更新导致以各版本软件开发工具包(softwaredevelopmentkit,SDK)也经常更新，以此为基础开发的应用程序也更新极快，恶意软件样本也随时间的进化速度较快，所以概念漂移对检测模型的影响较大。API更新速度太快导致恶意检测软件随着SDK更新迭代表现更差。EveDroid \cite{lei2019evedroid}和DroidSpan \cite{droidSpan}尝试找到一些可区分的且复杂的特征，从而可以建立更加具有可持续性使用的模型。Transcend \cite{transcending2022}提出使用统计技术在模型性能开始急剧下降之前检测概念漂移。InterDroid\cite{interDroid}提出的恶意软件检测方法引入了域自适应的方法，在已有数据和模型的基础上，采用并基于联合分布适配(joint distribution adaptation,JDA)算法进行特征迁移，以少量无标注的新Android软件为代价缓解了概念漂移问题。然而其漂移测试的时间只有两年，并没有数据佐证他们的模型可以对抗恶意软件更多年份的进化。
Concept drift is a common phenomenon in machine learning where the statistical properties of samples change over time. In machine learning-based methods, concept drift refers to the phenomenon where the statistical characteristics of test samples change in unpredictable ways over time, generally resulting in decreased model detection accuracy. Unlike traditional feature-based classification tasks, such as cat and dog classification, where the features of cats and dogs are very stable and hardly change over time, operating systems like Android and Windows frequently update to significantly increase market share. This leads to frequent updates of software development kits (SDKs) for various versions, and applications developed on this basis, including malicious software samples, also evolve rapidly. Thus, concept drift significantly impacts detection models. The rapid pace of API updates causes malware detection software to perform worse with each SDK iteration. EveDroid \cite{lei2019evedroid} and DroidSpan \cite{droidSpan} attempt to identify distinguishable and complex features, thereby establishing more sustainable models. Transcend \cite{transcending2022} proposes using statistical techniques to detect concept drift before model performance starts to decline sharply. The malware detection method proposed by InterDroid\cite{interDroid} introduces domain adaptation techniques. Building on existing data and models, it utilizes the Joint Distribution Adaptation (JDA) algorithm for feature transfer, mitigating concept drift at the cost of a small amount of unlabeled new Android software. However, their drift test spanned only two years, and there is no data to support that their model can withstand the evolution of malware over longer periods.

\subsection{Malware Detection Image-based Methods:}
% *************************不要了****************
% The concept of malware visualization is to visualize malware as images that can be fed into various visualization architectures for classification. First proposed by Nataraj et al.(2011) \cite{first_malware_image} that targets representing the bytes of malware as pixels of an image. They transformed malware binary information into images and then extracted GIST features for classification. They used the K-nearest neighbor classification technique on the Malimg\cite{malimg_dataset} dataset. Kesav Kancherla et al.\cite{Kancherla_2016_2013} converted each malware executable into the image and then extracted their texture features. They calculated the similarity between these features using Support Vector Machine (SVM).
% ************************不要了****************
% 恶意软件可视化的概念是将恶意软件可视化为图像，这些图像可以输入到各种可视化体系结构中进行分类。Kancherla等\cite{kancherla2013imageVisualization}把可执行文件转换为字节图的灰度图像，然后提取低级特征，如基于强度和基于纹理的特征等，最后使用SVM进行分类。之后有许多类似的工作出现\cite{kancherla2016packer}, \cite{imcfn}, cite{Yuan2020ByteLevel}, \cite{minihash}, \cite{malwareOnTheBrain}。最新的基于图像检测恶意软件的方法是ResNeXt+ \cite{2024resNeXt}, 这个方法基于ResNeXt模型加入了不同的即插即用注意力机制，模型通过捕获恶意软件图像通道感知视场来训练其专注于恶意软件特征，并且能够提供比其他方法更有用和灵活的信息，该方法的结果表明注意力机制可以提高恶意软件检测和分类的准确率。
The concept of malware visualization involves transforming malware into images that can be fed into various visualization architectures for classification. Kancherla et al.\cite{kancherla2013imageVisualization} converted executable files into grayscale images of byte plots, then extracted low-level features such as intensity-based and texture-based features, ultimately classifying them using SVM. Subsequently, many similar works emerged \cite{kancherla2016packer}, \cite{imcfn}, \cite{Yuan2020ByteLevel}, \cite{minihash}, \cite{malwareOnTheBrain}. The latest image-based malware detection method is ResNeXt+ \cite{resNeXt+2024}, which builds on the ResNeXt model by incorporating various plug-and-play attention mechanisms. The model trains to focus on malware features by capturing the perceptual field of malware image channels and provides more useful and flexible information than other methods. The results demonstrate that attention mechanisms can enhance the accuracy of malware detection and classification.

\subsection{SAM And Prompt Method}
\begin{figure}[!t]
\vspace{-1.0em}
\centering
\includegraphics[scale=0.42]{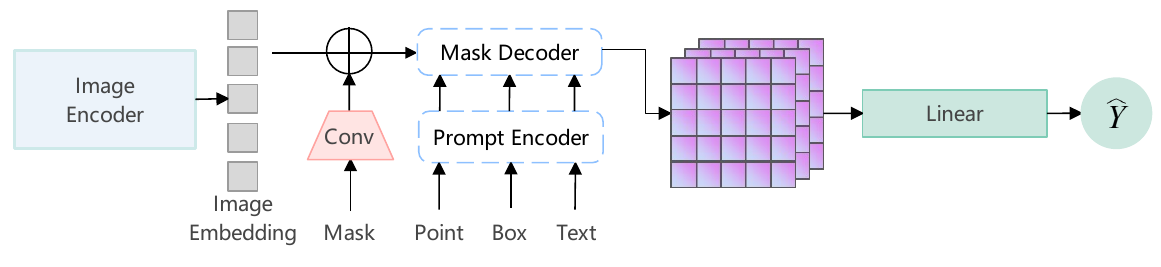}
\caption{The overflow of Segment Anything Model}
\label{fig_ordinary_sam}
\vspace{-1.0em}
\end{figure}

% Segment Anything Model(SAM)原本旨在构建一个基础模型用于图像分割\cite{2023SAM}。这个方法是利用通过提示工程解决下游任务中的分割问题，从而在新的数据分布上实现强大的泛化能力。如图\ref{fig_ordinary_sam} 所示，SAM模型大致分为三个部分：Image encoder，prompt encoder和mask decoder。图像通过一个庞大的以MAE Vit为基础架构的Image Encoder生成一系列的image embeddings，这些embeddings会输入到后续的模块中用于学习。SAM的prompt encoder接收两种prompt：dense prompt（例如mask）和sparse prompt（包括point和box）。SAM的mask decoder block maps the embeddings from image encoder and prompt encoder to a mask. The mask decoder is designed based on a Transformer followed by a prediction head. The image embeddings will be upsampled and then classified at each image location.
% SAM（Segment Anything Model）最初旨在构建一个基础模型，用于图像分割。这种方法的核心目标是利用预训练模型，通过提示工程解决分割问题，从而在新的数据分布上展现强大的泛化能力。SAM可以基于输入提示（如点或框）生成高质量的对象掩模，并可用于为图像中的所有对象生成掩模。SAM模型大致分为三个部分：图像编码器、提示编码器和掩码解码器。图像首先通过一个基于Masked Autoencoders(MAE) Vit的大型架构的图像编码器生成一系列图像嵌入，然后这些嵌入将用于后续模块的学习。SAM的提示编码器接收两种提示：dense prompt（例如mask）和sparse prompt（包括point和box）。SAM的掩码解码器块将来自图像编码器和提示编码器的嵌入映射到一个掩码上。掩码解码器基于Transformer设计，并带有一个预测头。图像嵌入将被上采样，然后在每个图像位置进行分类。
The Segment Anything Model (SAM) was originally designed to establish a foundational model for image segmentation \cite{2023SAM}. This method leverages pre-trained models through prompt engineering to address segmentation tasks, thus exhibiting strong generalization capabilities on new data distributions. As illustrated in Figure \ref{fig_ordinary_sam}, the SAM model is broadly divided into three parts: Image Encoder, Prompt Encoder, and Mask Decoder. Images are processed through a substantial architecture based on Masked Autoencoders(MAE) Vit to produce a series of image embeddings, which are then fed into subsequent modules for learning. SAM's Prompt Encoder receives two types of prompts: dense prompts (such as masks) and sparse prompts (including points and boxes). The Mask Decoder block in SAM maps the embeddings from the Image Encoder and Prompt Encoder to a mask. The Mask Decoder is designed based on a Transformer, followed by a prediction head. The image embeddings are then upsampled and classified at each image location.

\section{METHODOLOGY}\label{methodology}
In this section, we discuss technique details of the four phases in PromptSAM+. As shown in Fig 1. PromptSAM+ goes through four phases: 1) Covertor, 2) PromptSAM Module Get Image Feature, and 3) Image-based detection. To help with security analysis, PromptSAM+ also makes efforts to classify malware family in the last phase. We focus on the Covertor and PromptSAM sections.

\begin{figure*}[!t]
\centering
\includegraphics[scale=0.52]{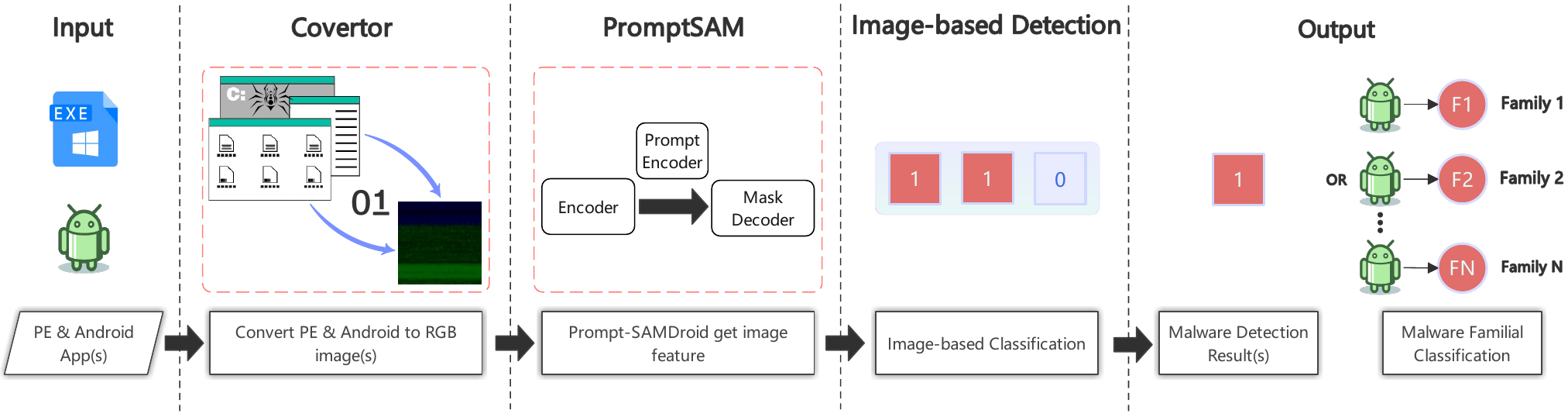}
\caption{The overflow of `PromptSAM+' System}
\label{fig_overflow}
\end{figure*}

First, we disassemble APK(Android Application Package) and extract $\mathbf{DEX}$(Dalvik Executable) files, that files are bytecode, after which we convert the $\mathbf{DEX}$ to the corresponding image by statical analyzing the dex file. Then, the model leverages Prompt-SAM techniques to perform automatic learning on image-structured data to get malicious features. 

\subsection{Covertor}\label{Convertor}
% 对于Windows PE文件，我们采取现有公开的数据集Malimg以及MaleVis，这些数据来源将PE读取二进制转换成图像,详情见Malevis网站\footnote{https://web.cs.hacettepe.edu.tr/~selman/malevis/}和Malimg\cite{first_malware_image}。

% 对于APK文件，APK文件包含以DEX文件形式存在的可执行字节码文件，其中包含用于运行的应用程序的已编译代码。如图\ref{fig_dex_structure2Image}左半部分所示的DEX文件结构主要由三部分组成：包括了(1)Header, (2)Ids, (3)Data. 其中Header section存储DEX文件的信息以及数据在文件中的偏移量，Ids section存储文件引用的所有字段、类型、字符串、方法、call_site标识符的索引列表, Data section存包含Ids section中所有列表的数据。Dalvik Executable规范将单个DEX文件中可以引用的方法总数限制在65,536个，包括Android框架方法、库方法和自己代码中的方法。要超过这个限制，需要配置应用程序构建过程以生成多个DEX文件，这个技术被称为多重DEX配置. 

% 因此，在一些APK文件中可能有多个DEX文件。对于我们提出的特征提取方法，Prompt-SAMDroid一张图像仅印射一个标签，无法做到多对一的印射。一个DEX只能对应一张图像，只能处理一个DEX文件，所以无法处理存在多个DEX文件的情况。我们如果仅仅处理第一个Dex而忽略了其他DEX文件，导致整个APK的信息将大量丢失，使得模型的检测效果不理想，这显然是不合理的。为了解决这个问题，我们需要集成多个DEX文件。根据对含有多个不同DEX文件的APK文件进行分析，我们观察到即使每个DEX文件的内容不同，但是DEX文件之间的section结构仍然相同。因此，我们可以将不同DEX中具有结构相同的片段分开，然后拼接不同DEX之间结构相同部分的片段，在文件可视化的过程中，拼接段可以直接作为新段使用,最终将多个DEX文件组成一个完整文件。这个过程如图Fig. 4所示

% 具体来说，对于每个APK文件我们使用apktool进行解压得到.dex文件，dex文件是APK中包含了源码的文件。我们使用apktool工具从一个APK文件中提取所有的DEX文件，然后我们读取每个DEX文件的bytesteam, 通过androguad工具分别找到DEX文件的Header，index，Data区块开始和结束部分, 将它们组合起来成为一个完整的DEX bytestream文件。 如图x所示，将完整的DEX文件转换为由8位无符号整数组成的一维数组。数组中的每一项都在[0,255]范围内，其中0代表黑色像素，255代表白色像素。然后，我们使用标准线性绘图将每个1D字节数组转换为二维(2D)数组，其中图像的宽度是固定的，高度也是固定的，并使用Pillow库中的标准Lanczos滤波器。最后，我们根据每个字节的用途为其分配颜色，在原始字节码的基础上添加一层语义信息，整个转换过程伪代码如算法2所示。我们将header section和Ids section以及data section根据Gamut进行分配RGB通道，可以得到由Dex转换相对应的图像数据，正如图片2右边部分的Dex File Image所示. 

\begin{figure}[!t]
\vspace{-1.0em}
\centering
\includegraphics[scale=0.44]{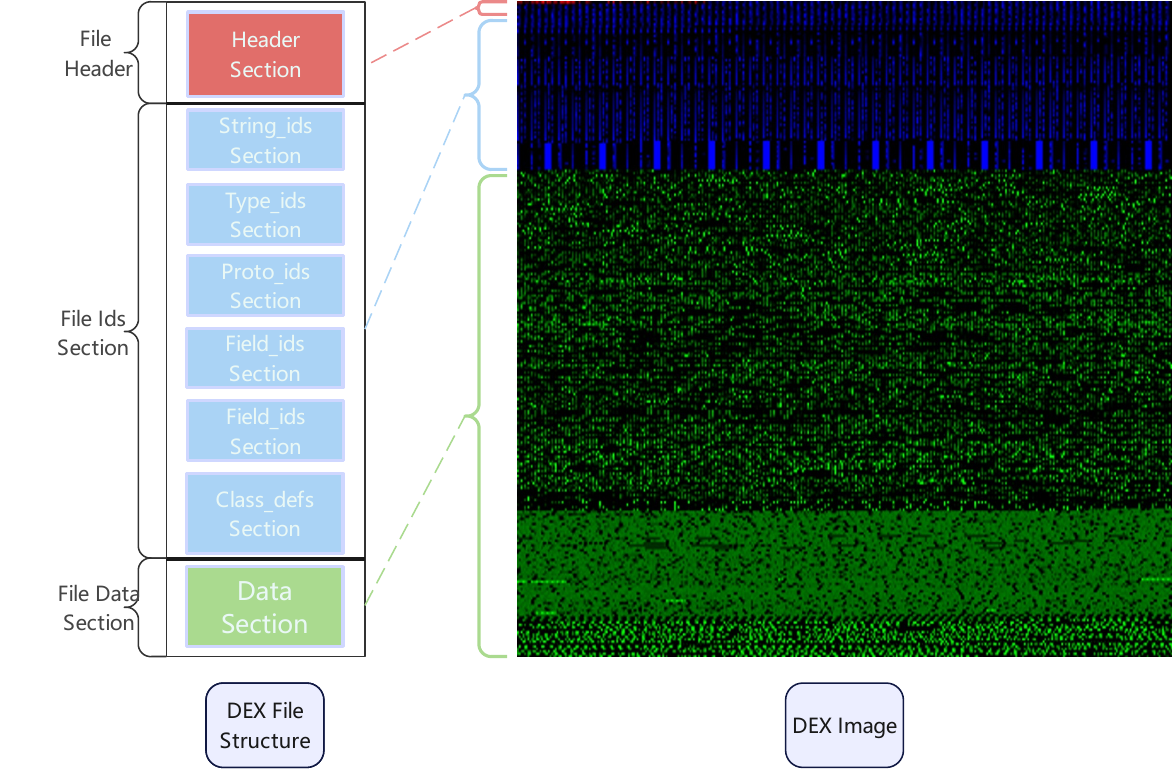}
\caption{The Overview Of Dex to Image. Left: Android DEX file structure, There are three main parts:(1)Header,2(Ids),(3)Data. Right: binary image representation of the Dex file.}
\label{fig_dex_structure2Image}
\vspace{-2.0em}
\end{figure}
\begin{figure}[!t]
\centering
\includegraphics[scale=0.3]{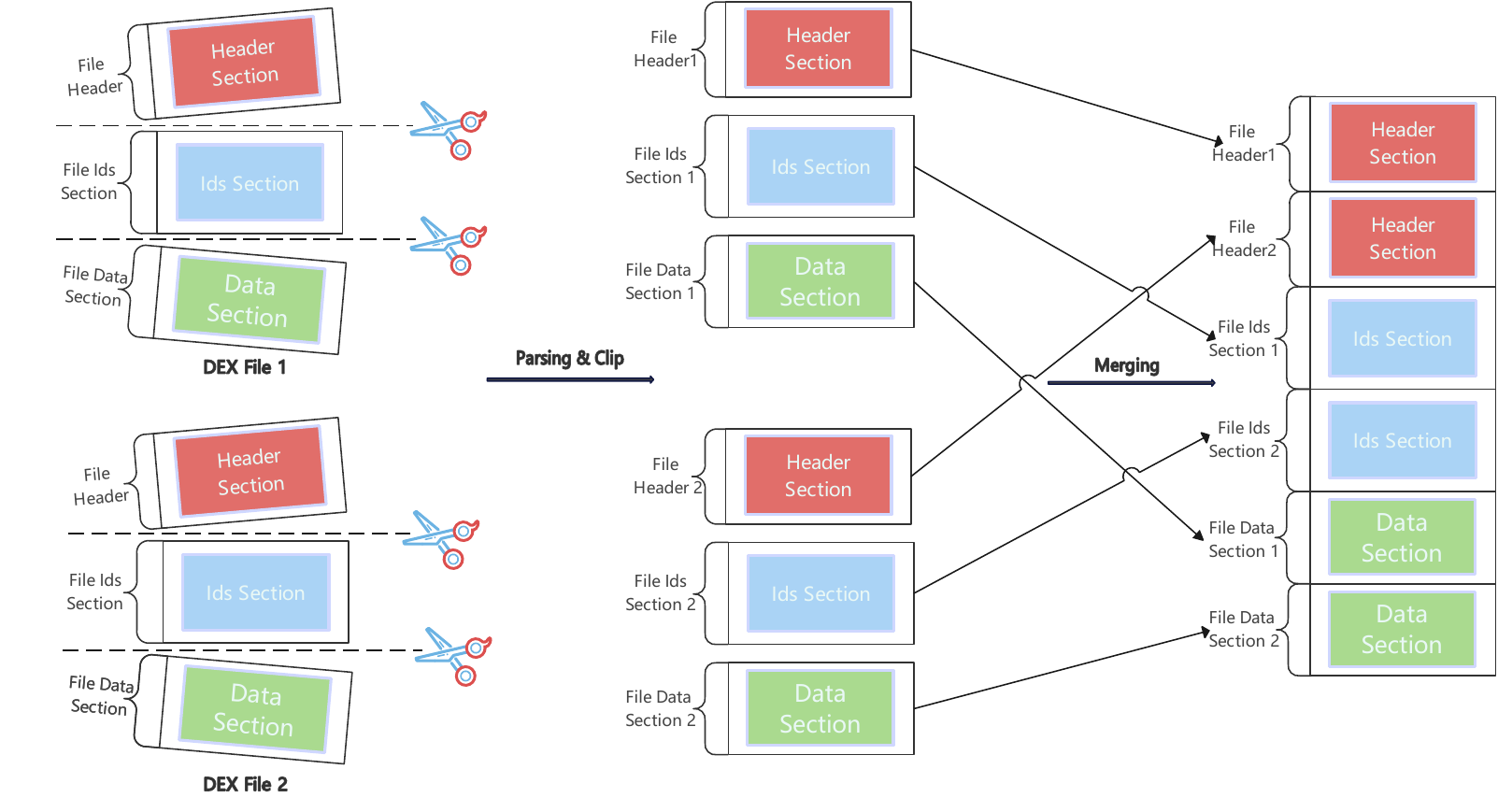}
\caption{Multiple dex files are decomposed, combined, and merged into one complete dex file.}
\label{fig_multi_dex_merger}
\vspace{-1.0em}
\end{figure}

For Windows PE files, we utilize publicly available datasets such as Malevis and Malimg, which convert PE read binaries into images. Details can be found on the MaleVis website\footnote{https://web.cs.hacettepe.edu.tr/~selman/malevis/} and in Malimg\cite{first_malware_image}.
 For APK files, APK files contain executable bytecode files in the form of $\mathbf{DEX}$ files, which contain the compiled code used to run app. The structure of a $\mathbf{DEX}$ file, as shown in the left half of Fig. \ref{fig_dex_structure2Image} under 'DEX File Structure', primarily consists of three parts: (1) Header Section, (2) Ids Section, and (3) Data Section. The Header section stores information about the $\mathbf{DEX}$ file and the data offsets within the file. The Ids section holds the index lists of all fields, types, strings, methods, and call site identifiers referenced in the file. The Data section contains the data for all the lists mentioned in the Ids section. The Dalvik Executable\cite{dalvik} specification limits the total number of methods that can be referenced within a single DEX file to 65,536, including Android framework methods, library methods, and methods in our own code. Getting past this limit requires that you configure your app build process to generate more than one DEX file, this technology is known as a multidex configuration.\cite{multidex}. Therefore, some APK files may contain multiple DEX files. For the feature extraction method we proposed, PromptSAM, one image maps to only one label, making a many-to-one mapping impossible. A single DEX can only correspond to one image and can only process one DEX file, so it's incapable of handling situations with multiple DEX files. If we only process the first Dex while ignoring the others, it would result in a significant loss of information from the entire APK, leading to suboptimal detection performance of the model, which is clearly unreasonable. To address this issue, we need to integrate multiple DEX files. Upon analyzing APK files containing several different $\mathbf{DEX}$ files, we observed that even though the contents of each DEX file differ, the section structure between $\mathbf{DEX}$ files remains the same. Therefore, we can separate segments with the same structure from different DEXs, and then concatenate the segments of structurally identical parts from different DEXs. In the process of file visualization, the concatenated segments can be directly used as new segments, ultimately combining multiple DEX files into a complete file. This process is illustrated in Fig. \ref{fig_multi_dex_merger}. More specifically, We use the apktool\cite{apktool} to extract all DEX files from an APK file, then we read the bytestream of each DEX file. Using the androguard\cite{androguard} tool, we identify the start and end parts of the Header, Index, and Data sections of the $\mathbf{DEX}$ files respectively. These sections are then combined to form a complete DEX bytestream file. Subsequently, the extracted DEX files are converted into a one-dimensional(1D) array composed of 8-bit unsigned integers. Each entry in the array is in the range [0, 255] where 0 corresponds to a black pixel and 255 a white pixel. We convert each 1D byte array into a two-dimensional(2D) array using standard linear plotting where the width of the image is fixed and the height is also fixed and using a standard Lanczos filter from the Pillow library\footnote{ https://github.com/python-pillow/Pillow}. Finally, we assign colors to each byte based on its purpose, adding a layer of semantic information on top of the original bytecode. We assign the $header$ section and $Ids$ section and $Data$ section to RGB channels according to Gamut\cite{gamut2017} algorithm, as shown in the DEX File Image in the right section of Fig. \ref{fig_dex_structure2Image}.

% \begin{figure}[!t]
% \centering
% \includegraphics[scale=0.39]{fig_dex2image_processing.pdf}
% \caption{The Processing of Dex file.}
% \label{dex2image_processing}
% \end{figure}

\begin{algorithm}[h]
\label{algorithm1}
\scriptsize
\renewcommand{\arraystretch}{1}
\caption{Algorithm describing 2D array RBG image generation from an APK.}
\label{alg:algorithm1}
\KwIn{APK File}
\KwOut{2D RGB image $\mathcal{G}$}  
\BlankLine
Initialized apktool and androguard tools \par
   $fixedWidth = 1024,512,256$\;
   $dexNum = apktool(APK).getDexNum()$\;
   $byteStream,dexHeader,dexIndex,dexData \leftarrow \emptyset$\; 
   \tcc{Walk through all $\mathbf{DEX}$ files that were disassembled from the APK}
    \For{$i=1; dexFile_i$ \KwTo $i=rang(dexNum)$}{
        $stream = dexFile_i.toByteStream()$\;
        $dexHeader = stream\_header\_start\ \KwTo \ stream\_header\_off$\;
        $dexIndex = stream\_index\_start\ \KwTo \ stream\_index\_off$\;
        $dexData = stream\_data\_start\ \KwTo \ steam\_data\_off$\;
    }
  \tcc{Merge all bytestreams into one complete $\mathbf{DEX}$ file}
  $byteStream = Concat(dexHeader, dexIndex, dexData)$\;
  \tcc{Reshape 1D bytestream array to 2D, the width and height equal fixed width}
  $create\_linear\_image(byteStream$\; 
  \For{width \KwTo fixedWidth}{
    $img.resize\_to\_size(height=width, width=width)$\;
    $img.save()$\;\tcc{save image $\mathcal{G}$}
  }
\end{algorithm}

\subsection{PromptSAM}
% 这里详细的写:
% 1. model的改进和原SAM的不同点
% 2. model内部具体的细节和全过程计算公式，loss function, target function, 优化目标等。
% 3. model的创新点是什么，我们的优势在哪，为什么有这样的优势的\textbf{证明}
% 4. model的overview部分的图，以及内部的细节图
% 5. model为啥work的原因
% 以上内容要占至少2页的内容(算上插图)

% 1. SAM的model结构
% 2. 

\subsubsection{PromptSAM Overview}
\begin{figure}[!t]
\vspace{-2.0em}
\centering
\includegraphics[scale=0.30]{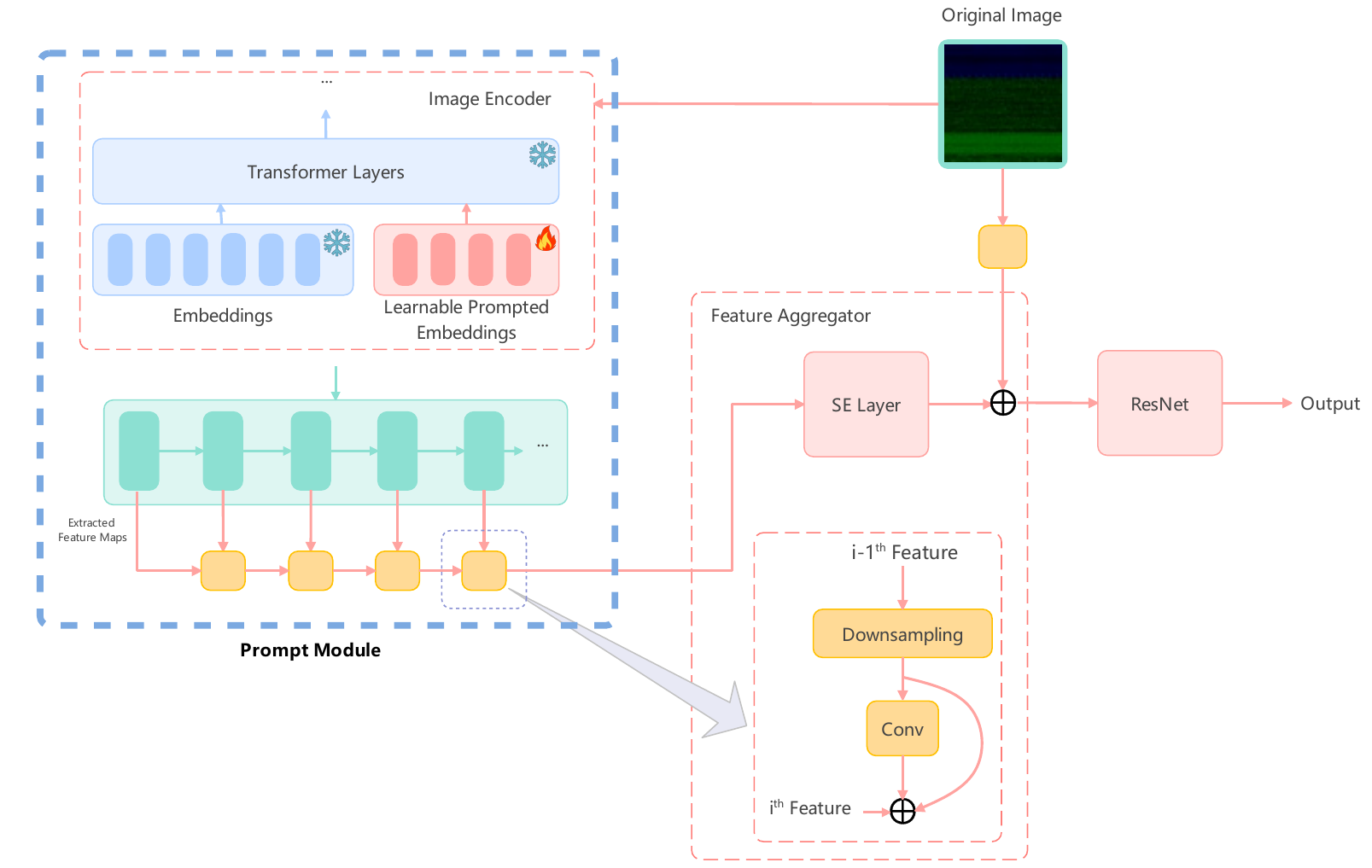}
\caption{The Overflow Of The 'PromptSAM+' Structure}
\label{fig_PromptSAM_module}
\vspace{-1.0em}
\end{figure}

% 图\ref{fig_PromptSAM_module}是我们提出的模型结构。本文方法的目的是在预训练SAM模型学习到的图像分割领域知识的指导下，学习跨领域的对恶意软件的分类知识。恶意软件图像通过两条路径输入，一个分支是Prompt模块（如图中蓝色框部分所示），另一条是降采样与pad后的原图。两条途径得到的结果会在进入主干网络之前融合。我们定义输入图片为i，Prompt模块为$\psy(x)$，降采样操作为$\phi(i)$，那么两条分支的结果可以形式化定义为$y=\psy(i)+\phi(i)$。降采样的目的是为了保证两个分支的维度匹配。
Figure \ref{fig_PromptSAM_module} presents the model structure proposed in this paper. The objective of this method is to learn cross-domain knowledge for classifying malware, guided by the domain knowledge of image segmentation acquired by the pre-trained SAM model. Similar to ResNet, we use a residual structure. Malware images are input through two pathways: one branch is the `Prompt Module' (as shown in the blue boxed section in the figure), and the other is the $Downsampling$ and padded original image. If the input image is $i$, the `Prompt Module' is $\rho(i)$  and the  $Downsampling$ operation is $\delta(i)$, then the results of the two branches can be formally defined as y=$\rho(i)$ + $\delta(i)$. The purpose of  $Downsampling$ is to ensure dimensional matching between the two branches.

% 这意味着我们为常见的分类网络制作了一个可插入模块，我们给这个可插入模块取名为'PromptSAM+'。该模块的输出和原图像一起输入到后续的（主干网络）分类模块中。其中Prompt模块是基于SAM进行改进的，它分为三个子模块：（1）对SAM的Image Encoder的ViT结构增加Learnable Prompted Embeddings的模块。（2）对Image Encoder提取feature map的子模块。（3）聚合第（2）部分中获得的feature模块的feature aggregator模块。下面我们将详细介绍各个子模块。
This means that we have created an insertable module for common classification networks, which we have named `PromptSAM+'. The output of this module, along with the original image, is input into the subsequent (backbone) classification module. The `Prompt Module' is based on improvements to SAM and is divided into three sub-modules: (1) a module that adds \textit{Learnable Prompted Embeddings} to the ViT structure of SAM's Image Encoder, (2) a sub-module that extracts feature maps from the Image Encoder, and (3) a feature aggregator module that aggregates the features obtained from part (2). Below, we will detail each part.

% \textbf{Learnable Prompted Embeddings in Image encoder：}在Prompt模块中，我们利用的是pretrained SAM模型，不需要重新训练，即整个SAM模型基本是被frozen的（但并非完全frozen）。给定一张特定恶意软件生成的图像，首先输入SAM的image encoder得到这部分的输出image embeddings。但为了保证SAM预训练模型对恶意软件分类任务的适应性，我们冻结SAM的大多数参数，将一些Learnable Prompted Embeddingds注入到SAM image encoder的块中，以提升整体结构的性能。（如图中的Learnable Prompted Embeddings模块所示）。The i-th(i=0,1,2,...) learnable promted embeddingds are inserted after the origin embeddings of the SAM ViT structure。 SAM的image encoder包含$n$个attention block，我们抽取每一个attention模块的输出（如图中extracted feature maps所示），这些feature maps将在后续被使用。
\textbf{Learnable Prompted Embeddings: } In the `Prompt module', we utilize the pretrained SAM model, which does not require retraining; thus, the entire SAM model is essentially frozen. Given a specific image generated by malware, it is first input into SAM's image encoder to obtain initial image embeddings. However, to ensure the adaptability of the SAM pretrained model to the malware classification task, we freeze most parameters of SAM and inject some \textit{Learnable Prompted Embeddings} into the image encoder blocks of SAM to enhance the performance of the entire structure (as shown in the \textit{Learnable Prompted Embeddings} in the figure\ref{fig_PromptSAM_module}). The i-th (i=0,1,2,...n) \textit{Learnable Prompted Embeddings} are inserted after the embeddings of the SAM ViT structure. SAM's image encoder contains $n$ attention blocks, and we extract the output of each attention module (as shown in the extracted feature maps in the figure), which will be used subsequently.

% \textbf{Feature Aggregator}：在SAM的image encoder中抽出的feature被aggregator聚合。聚合器聚合的结果由一个SE layer再次捕获信息。
\textbf{Feature Aggregator:} In the SAM image encoder, the extracted features are aggregated by an aggregator module. These features are input into the feature aggregator module. The results aggregated by the aggregator are further captured by an SE layer.

% \textbf{Classifier}： SE layer的输出被送入Detector $D$（PromptSAM+的head部分）进行最终的恶意软件分类。
\textbf{Classifier:} The output of the SE layer is fed into the Detector $D$ (the head part of PromptSAM+) for the final classification of malware.

\subsubsection{Image Encoder Prompting Block}
%  这里解释一下learnable embeddings为什么要这么做
%  与经典ViT结构中图片被映射为embeddings的方法相同，输入图像$I_mal$被划分为$n$个patch以生成$n$个embeddings，然后输入到SAM的Image Encoder的ViT。我们在SAM的image encoder的ViT中的第一个Transformer层中注入了一些Learnable Prompted Embeddings $L$.我们将原有SAM模型的embeddings $e^i$的生成过程表示为$Em$, 如公式\ref{em_def}：
We inject some \textit{Learnable Prompted Embeddings} into the first Transformer layer $L$ of the image encoder of SAM. The input image $I_{mal}$ is divided into $n$ patches that are input into the Transformer to generate $n$ embeddings. We denote the process of generating these embeddings $e^i$ as $Em$, as equation\ref{em_def}:
\begin{equation}
e^i = Em(I_{mal})  \qquad e^i\in \mathbb{R}^d\times d,i=1,2,...,m.
\label{em_def}
\end{equation}

% 我们对$L$产生的patched embeddings增添$m$个d*d维的Learnable Prompted Embeddings，冻结住image Encoder的其他embeddings，保证只对Learnable Prompted embeddings训练, 计算公式为\ref{xp_def} \ref{y_def}：
We augment the patched embeddings produced by $L$ with $m$ \textit{Learnable Prompted Embeddings} of dimension $d \times d$. We freeze the other embeddings of the image encoder, ensuring that only the \textit{Learnable Prompted Embeddings} are trained, as equation \ref{xp_def} \ref{y_def}:

\begin{equation}
\label{xp_def}
[x_i,x_p]=\mathcal{L}_i ([e^i,e^p])
\end{equation}

\begin{equation}
\label{y_def}
y = Head([x_i,x_p])
\end{equation}

% 其中$e^p$是我们插入的Learnable Prompted Embeddings，$\mathcal{L}_i为整个映射函数，其中$e^i$部分是冻结的，$e^p$部分的参数是可以更新的。$Head([x_{i},x_{p}])$表示分类头的过程，$y$代表正向传播过程中整个image encoder的输出，即image embedding。% 由此，我们就构建出了Prompt Module的第一部分Image Embedding。
The $e^p$ represents the \textit{Learnable Prompted Embeddings} that we introduce, while $\mathcal{L}_i$ is the mapping function for the entire procedure. Within this setup, the part $e^i$ parameters are frozen, while the parameters of the $e^p$ part are updatable. $Head([x_{i},x_{p}])$  represents the process of sorting headers. $y$ represents the output of the entire image encoder during the forward propagation process, namely the image embedding. Consequently, this forms the first part of the module.

\subsubsection{Feature extractor}
% The feature extractor block consists of a group of Convolution layers to capture features from the image encoder blocks of SAM(特征提取器block由一组卷积层组成，用于从SAM的image encoder block中捕获特征). 我们只关注SAM的image encoder，SAM的image encoder模块的attention层含有大量的信息，我们抽取每次经过attention模块的feature map。这个过程如公式\ref{extract_feat}所示。其中\varepsilon_enc表示提取image encoder层信息的这个操作。
The feature extractor block consists of a group of convolution layers that capture features from the image encoder blocks of SAM. Our focus is solely on SAM's image encoder, whose attention layers contain a wealth of information. We extract the feature map each time it passes through the attention module. This process is shown in equation \ref{extract_feat}. Where $\varepsilon_enc$ represents this operation to extract information from the image encoder layer
\begin{equation}
\label{extract_feat}
f_i=\varepsilon_{enc} (i)
\end{equation}

% 值得注意的是，我们上一步插入的Learnable Promped Embeddings也跟随image encoder计算，参与image embeddings的输出。在预训练模型中，如前文所述，SAM的image encoder模块含有SAM对图片提取的隐藏信息，（值得注意的是，image encoder的feature部分也会含有prompt信息，也就是语义信息）。为了证明预训练视觉模型对语义的提取也会下游的恶意软件任务产生协同作用，我们从image encoder中抽取部分feature。为了获得更多的SAM的prompt信息，我们采用的是dense prompt预训练模型。
It should be noted that the \textit{Learnable Promped Embeddings} obtained in the previous step are also processed with the image encoder, contributing to the output of the image embeddings. In the pretrained model, as previously mentioned, SAM's image encoder module contains hidden information extracted from the images (it is important to note that the feature part of the image encoder also contains prompt information, i.e., semantic information). To demonstrate that the extraction of semantics by the pretrained visual model also synergistically affects downstream malware detection tasks, we extract some features from the image encoder. To gain more of SAM's prompt information, we utilize a dense prompt pretrained model.

% \begin{equation}
% f_u = \varepsilon_{dec}(y)
% \end{equation}

\subsubsection{Feature aggregator}
\begin{figure}[!t]
\centering
\includegraphics[scale=0.3]{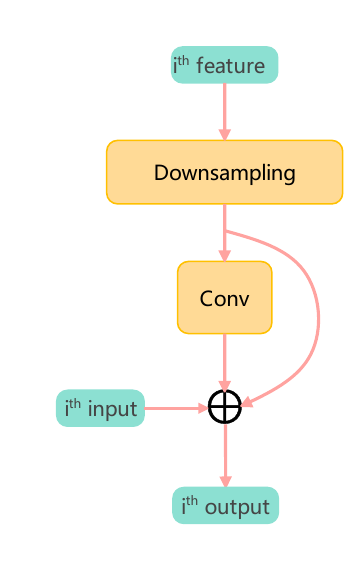}
\caption{The Feature Aggregator}
\label{fig_feature_agg}
\vspace{-1.0em}
\end{figure}
% 从SAM的image encoder中提取出的feature maps会进入我们设计的feature aggregator to aggregate, followed by a SE attention block to integrate and process the overall information. feature aggregator由几个相同的子模块组成的，每个子模块有两个输入，分别是当前image encoder block抽出来的feature $f_i$和经过上一个模块处理后output。也就是说，当前子模块的输出是下一个子模块的输入。子模块接收到输入feature之后会对其进行downsample、卷积和残差处理，如图\ref{fig_feature_agg}所示，类似残差结构，$f_i$是当前层抽出的feature，$\mathcal{F}(f_i)$是子模块对抽出的feature进行处理的过程，\mathcal{D}(f_i)是降采样,\psi(f_i)是当前子模块的输出，\psi(f_{(i-1)})是上个子模块的输出，也是当前子模块的输入之一（除第一个子模块外）。过程如公式\ref{aggregator_def} \ref{res_def}所示。其中$\mathcal{D}(f_i)$表示对当前的feature $f_i$进行降采样，$\psi()$是残差操作。
The feature maps extracted from SAM's image encoder are fed into our designed feature aggregator to aggregate, followed by an SE attention block to integrate and process the overall information. The feature aggregator consists of several identical sub-modules, each with two inputs: the feature $f_i$ extracted from the current image encoder block and the output processed by the previous module. This means that the output of the current sub-module serves as the input for the next sub-module. Upon receiving the input features, each sub-module performs downsampling, convolution, and residual processing, as shown in Figure \ref{fig_feature_agg}. This setup is similar to a residual structure, where $f_i$ is the feature extracted from the current layer. $\mathcal{F}(f_i)$ represents the process by which the sub-module handles the extracted feature, $\mathcal{D}(f_i)$ is the downsampling, and $\psi(f_{(i)})$ is the output of the current sub-module. $\psi(f_{(i-1)})$ is the output from the previous sub-module and also one of the inputs for the current sub-module (except for the first sub-module). The process is illustrated in equations \ref{aggregator_def} and \ref{res_def}. where, $\mathcal{D}(f_i)$ denotes the downsampling of the current feature $f_i$, and $\psi()$ represents the residual operation.

\begin{equation}
\label{aggregator_def}
\psi(f_i) = \psi(f_{(i-1)}) + \mathcal{F}(f_i)
\end{equation}

\begin{equation}
\label{res_def}
 \mathcal{F}(f_i)=\mathcal{D}(f_i)+\mathcal{D}(Conv(f_i))
\end{equation}

\subsubsection{Our Strength}
% PromptSAM+的head网络（也就是Detector $D$）接收的输入一部分来自Prompt模块，另一部分来自原图像，而非完全是SAM预训练模型中提取到的信息。因此我们的PromptSAM有以下优势：
% 1.我们的模型尝试用视觉领域的预训练模型对恶意软件检测进行分类，尽管数据集的分布相差很大，但是该预训练模型仍旧对任务有帮助。值得注意的是，PromptSAM+ 网络的detector部分并没有见过原图像，而是只接受了经过feature aggregator处理过的SAM模型的feature map。也就是说，我们的模型研究了SAM对于下游任务的丰富的语义信息的作用。
% 2. SAM预训练模型使得我们的Prompt-SAMDroid网络在没有降低检测能力的情况下，获得了良好的抗漂移能力与家族分类能力。
% 3.我们的模型收敛速度极快，效果较好，这同样证明了SAM预训练模型的协同作用。
The head network of PromptSAM+ (also known as Detector $D$) receives inputs partly from the \textit{Prompt module} and partly from the original image, but entirely based on the information extracted from the SAM pretrained model. Therefore, our PromptSAM+ has the following advantages:

\begin{enumerate}
\item {Our model attempts to classify malware detection using a pretrained model from the visual domain. Despite the significant differences in dataset distribution, the pretrained model still aids the task. Notably, the detector part of the PromptSAM+ network has never seen the original image; it only receives the feature map processed by the feature aggregator from the SAM model. This means that our model explores the rich semantic information of SAM for downstream tasks.}

\item {The SAM pretrained model enables our PromptSAM+ network to achieve good drift resistance and family classification abilities without compromising detection capabilities.}

\item {Our model converges rapidly and performs well, which also demonstrates the synergistic effect of the SAM pretrained model.}
\end{enumerate}

\section{EXPERIMENTS AND RESULTS}\label{expriments_and_result}
\subsection{ Dataset Description And Experimental Setup}

\begin{table}[!t]
\tiny
\renewcommand{\arraystretch}{1}
\setlength{\tabcolsep}{5pt}
\caption{DATASETS USED IN THIS WORK:\label{dataset_description}}
\centering
\begin{tabular}{cccccp{4cm}}
\hline
\textbf{Datasets} & \textbf{Family} & \textbf{OS} & \textbf{Quantity} & \textbf{Time}\\
\hline
MaleVis & \textbf{25}  & Windows   & 14226       &       -        \\ 
Malimg  & \textbf{25}  & Windows   & 9339        &       -        \\
Prompt-Family & \textbf{24} & Android &  27294   &       -        \\
Prompt-Time   & \textbf{2}  & Android &  105000  &   2015-2021    \\
\specialrule{0.05em}{2pt}{2pt}
\textbf{total} &            &         &  155,859  &             \\ 
\specialrule{0.05em}{2pt}{2pt}
\end{tabular}
\end{table}

\begin{table}[!t]
\tiny
\renewcommand{\arraystretch}{1}
\setlength{\tabcolsep}{5pt}
\caption{DESCRIPTION OF PROMPT-TIME DATASET \label{prompt_family_dataset_description}}
\centering
\begin{tabular}{cccccp{1.5cm}}
\hline
\textbf{Year} & \textbf{Malware (M)} & \textbf{Benign (B)} & \textbf{M+B} & \textbf{M/(M+B)} \\
\hline
2015 $\longrightarrow$ 2021 & 3000  & 12000  &  15000  &  20\%         \\ 
% 2016 & 3000  & 12000  &  15000  &  20\%         \\
% 2017 & 3000  & 12000  &  15000  &  20\%         \\
% 2018 & 3000  & 12000  &  15000  &  20\%         \\
% 2019 & 3000  & 12000  &  15000  &  20\%         \\
% 2020 & 3000  & 12000  &  15000  &  20\%         \\
% 2021 & 3000  & 12000  &  15000  &  20\%         \\
\specialrule{0.05em}{2pt}{2pt}
\textbf{total} &      &         & 105000  &     \\
\specialrule{0.05em}{2pt}{2pt}
\end{tabular}
\vspace{-2.0em}
\end{table}
% 我们使用两个基准恶意软件数据集以及自己收集整理的具有时间标记和家族标记的数据集用来评估所提出的恶意软件检测框架的性能。它们包括 Malimg\cite{first_malware_image}、MaleVis\cite{MaleVis}、Prompt-Family和 Prompt-Time。Prompt-Family 和 Prompt-Time 是由我们自行收集，分别来自 Androzoo \cite{androzoo}和MalNet \cite{malnet}项目，AndroZoo 是一个著名的 Android 应用程序集合，主要来自 Google Play、AnZhi和AppChina, 样本经过数十种不同的杀毒软件产品分析。它仍在更新中，目前包含 24,264,358个不同的良性软件和恶意软件样本。MalNet 是一个层次化的图像和图数据库，旨在帮助机器学习和安全研究人员开发检测恶意软件的方法。我们收集了 132,294 个 Android APK文件，其中 Prompt-Time 包含 105,000 个带有良性和恶意标签的样本，时间跨度从2015年-2021年共7年。Prompt-Family 包含 27,294个APK，其中既有家族标签又有类型标签，这些标签来自 Euphony \cite{Euphony}，这是一个先进的恶意软件标记系统，汇总并学习来自 VirusTotal \cite{virustotal}上多达 70 家杀毒软件供应商的标记结果。VirusTotal返回的标签可能是“恶意”或“良性”，由VirusTotal响应中的“detected”字段指示。这些标签的数量记为p。为了获得可靠的真实实验数据，同时也可以在现实世界中的数据分布，我们数据集遵循Zhu\cite{zhu2020measuring}等人的建议, 我们选择对于良性软件 p = 0, 对于恶意软件阈值为 20≤p≤29, 并且我们对每年的良性与恶意软件比例都控制在4:1, 这与现实世界中恶意软件与良性软件的比例也较为接近。Prompt-Family 和 Prompt-Time 被用来分析所提出的恶意软件检测框架的性能。Prompt-Family 数据集中包含的恶意软件家族和类型如表\ref{prompt_family_dataset_description}所描述。

We evaluated the performance of the proposed malware detection framework using two benchmark malware datasets along with our own dataset marked with timestamps and family labels. These include the Malimg\cite{first_malware_image}, MaleVis\cite{MaleVis}, Prompt-Family, and Prompt-Time datasets. Prompt-Family and Prompt-Time were self-collected, sourced from the AndroZoo \cite{androzoo} and MalNet \cite{malnet} projects respectively. AndroZoo is a renowned collection of Android applications mainly from Google Play, AnZhi, and AppChina, analyzed by dozens of antivirus products. It continues to be updated and currently contains 24,264,358 different benign and malicious software samples. MalNet is a hierarchical database of images and graphs designed to aid machine learning and security researchers in developing methods to detect malware. We have collected 132,294 Android APK files, where the Prompt-Time dataset includes 105,000 samples with benign and malicious labels spanning seven years from 2015 to 2021. Prompt-Family contains 27,294 APKs, each labeled with both family and type tags sourced from Euphony \cite{Euphony}, an advanced malware labeling system that aggregates and learns from the markings provided by up to 70 antivirus vendors on VirusTotal \cite{virustotal}. The labels returned by VirusTotal may be "malicious" or "benign," as indicated by the "detected" field in the VirusTotal response. The number of these labels is denoted as $p$. To obtain reliable experimental data and also reflect the distribution of data in the real world, our datasets follow the recommendations of Zhu et al.\cite{zhu2020measuring}, choosing a benign software threshold of $p=0$, and for malicious software a threshold of $20\leq p \leq 29$, while controlling the annual ratio of benign to malicious software at 4:1, which closely aligns with the real-world ratio of malware to benign software. Prompt-Family and Prompt-Time are used to analyze the performance of the proposed malware detection framework. The malware families and types included in the Prompt-Family dataset are described in Table \ref{prompt_family_dataset_description}.

% 我们使用PyTorch1.10 \cite{pytorch} 在Python中实现了我们的模型，包括其他所有复现的模型也都基于此版本。实验平台的硬件配置为Intel Xeon Silver 4215R (32) @ 4.000GHz，RTX 3090 24 GB * 2，128GB RAM，Ubuntu 20.04.4 LTS x86_64。
We implemented our model in Python using PyTorch 1.10 \cite{pytorch}, and all other replicated models are also based on this version. The experimental platform's hardware configuration includes an Intel Xeon Silver 4215R (32) @ 4.000GHz, RTX 3090 24 GB * 2, 128GB RAM, running on Ubuntu 20.04.4 LTS x86\_64.

\subsection{Experimental Results Metrics}
% 为了评估我们的方法，衡量PromptSAM+框架的有效性，我们利用一些广泛使用的指标，如准确率、精确率以及接收者操作特征曲线（ROC）来进行性能评估，如表\ref{metrics_description}所示。准确率是正确预测的恶意软件样本占所有恶意软件样本的比率。由于恶意软件数据集不平衡，我们还将召回率和F1分数作为评估指标。精确率表示阳性预测率，而召回率表示真阳性率。这些指标非常适合评估不平衡的深度学习分类问题。除此之外，在现实世界(real-time)恶意软件检测中，高误报率(FPR)可能导致用户对安全系统的不满和不信任，进而可能选择关闭安全功能，从而使系统更加脆弱。低漏报率(FNR)确保了恶意软件的高检测率，减少了恶意软件造成的潜在损害。

To evaluate our methods we conduct experiments by performing ten-fold cross-validations. Furthermore, to measure the effectiveness of `PromptSAM+', We leverage certain widely used Accuracy, F1-score, Recall, Precision, FPR, FNR for performance evaluation are shown in Table \ref{metrics_description}. The accuracy is the ratio of correctly predicted malware samples overall malware samples. Because the malware dataset is imbalanced we also take precision, recall, and F1-score as evaluation metrics. Precision means the positive predictive rate, while recall indicates the true positive rate These metrics are well-suited for evaluating the imbalanced deep learning classification problem.

% % 我们还检验了框架的效率。由于恶意软件分类在任何反病毒产品中都是一个时间敏感的组件，一个小的延迟可能会错过发现恶意软件进程的最佳机会，所以区分恶意软件样本的过程应该花费较短的时间。我们在测试过程中测量了通过的数据集的CPU时间，可以通过每个恶意软件样本的CPU处理时间来衡量，计算了每个恶意软件样本的平均MPE，这仅考虑特征提取和分类。\textit{MPE}定义如方程\eqref{MPE_def}, 其中\textit{cputime}表示计算资源消耗的时间，\textit{files_num}表示处理的文件数量。
% We also examine the efficiency of our framework. As Prompt-SAMDroid is a time-sensitive system, its efficiency, denoted by MPE, can be measured by the CPU processing time per malware sample that counts only feature extraction and classification. where \textit{cputime} indicates the time consumed by computing resources, and \textit{files\_num} indicates the number of processed files. \textit{MPE} defined as equation \eqref{MPE_def}
% \begin{equation}
% \label{MPE_def}
% MPE = \frac{cpu_time}{files\_num}
% \end{equation}

\begin{table}[!t]
\vspace{-2.0em}
\tiny
\renewcommand{\arraystretch}{1}
\caption{Descriptions of metrics used in our experiments:\label{metrics_description}}
\centering
\begin{tabular}{cccp{4pt}}
\hline
\textbf{Merics} & \textbf{Abbr} & \textbf{Definition} \\
\hline
True Positive  & \textbf{TP} & \makecell{The number of malware that \\ is correctly classified as malware}  \\
True Negative  & \textbf{TN} & \makecell{The number of benign that \\ is correctly classified as benign }     \\
False Positive & \textbf{FP} & \makecell{The number of benign \\ misclassified  as malware}  \\
False Negative & \textbf{FN} & \makecell{The number of malware \\ misclassified as benign }  \\
\hline
True Positive Rate  & \textbf{TPR} &  $\mathrm{TPR}={\frac{TP}{TP+FN}}$  \\ 
\specialrule{0em}{1pt}{1pt}
False Negative Rate & \textbf{FNR} &  $\mathrm{FNR}={\frac{FN}{TP+FN}}$  \\ 
\specialrule{0em}{1pt}{1pt}
True Negative Rate  & \textbf{TNR} &  $\mathrm{TNR}={\frac{TN}{TN+FP}}$  \\ 
\specialrule{0em}{1pt}{1pt}
False Positive Rate & \textbf{FPR} &  $\mathrm{FPR}={\frac{FP}{TN+FP}}$  \\
\hline
\specialrule{0em}{2pt}{2pt}
Accuracy  & \textbf{Acc} &  $\mathrm{Acc}={\frac{TP+TN}{TP+TN+FP+FN}}$ \\
\specialrule{0em}{2pt}{2pt}
Precision & \textbf{Pre} &  $\mathrm{Pre}={\frac{TP}{TP+FP}}$ \\ 
\specialrule{0em}{2pt}{2pt}
Recall    & \textbf{Rec} &  $\mathrm{Rec}={\frac{TP}{TP+FN}}$ \\ 
\specialrule{0em}{2pt}{2pt}
F1-Socre  & \textbf{F1}  &  $\mathrm{F1}={\frac{2TP}{2TP+FP+FN}}$ \\ 
% \specialrule{0em}{2pt}{2pt}
% ROC Area  & \textbf{AUC} &  Area under ROC curve \\
\specialrule{0.05em}{2pt}{2pt}
\end{tabular}
% \vspace{-2.0em}
\end{table}

\subsection{Malware Detection Ability And Results Analysis}
% 为了证明我们方法是可靠的并且可以对大规模筛查恶意软件的存在，利用我们提出的图像生成策略生成相对应的恶意软件图像，利用图像数据评估了模型对恶意软件的检测能力。具体来说，我们在Prompt-Time数据里选出年份为2015年用于作为训练集和测试集，我们使用了多个不同系列的深度学习视觉模型评估了我们的方法。我们将\ref{Convertor}章节部分转换的恶意软件图像用于恶意软件检测中。首先，为了探究输入的图像宽度对模型性能的影响，我们尝试将宽度为1024,512，256的图像输入模型中。以ResNet50模型作为示例，我们使用2015年的数据用于训练，测试集合为2015年-2021年的数据集，结果如表\ref{resnet50_for_256and1024}所示。我们发现当输入图像宽度为256，测试集时间同为2015年情况下，ResNet50模型的Acc性能表现为93.00%, 当输入图像宽度为1024, ResNet50模型的Acc性能表现为94.19%, 两者之间没有本质的差别。但是，当输入图像宽度为256或者图像宽度更小的情况时, 此时测试集为2015年之后的数据，模型在2021年的下降的幅度极大，到了2021年只有81.12%。相比之下输入宽度为1024的图像，测试集同为2015年之后的数据，此时模型的Acc性能并没有下降的很快，即使最低的情况还保留84.91%。我们认为发生这样的情况是因为在图像转换后，图像太小的宽度会导致部分数据的丢失，从而影响了精度，但是生成的图像宽度太大也会极大的增大模型的计算量，导致效率极低，因此我们为了平衡两者，之后的其他实验都选择以1024宽度作为输入的图像宽度, 避免了图像数据精度丢失也保证计算效率。
To demonstrate that our method is reliable and capable of screening malware on a large scale, we generated corresponding malware images using our proposed image generation strategy and assessed the model's malware detection capability using image data. Specifically, we selected the year 2015 from the Prompt-Time dataset to serve as both the training and testing sets. We evaluated our approach using multiple series of deep learning visual models. We used the malware images converted in section \ref{Convertor} for malware detection. Initially, to explore the impact of input image width on model performance, we attempted to input images of widths 1024, 512, and 256 into the models. Using the ResNet50 model as an example, we trained using data from 2015, with the test set spanning from 2015 to 2021, as shown in Table \ref{resnet50_for_256and1024}. We found that when the input image width was 256 and the test set was also from 2015, the accuracy (Acc) performance of the ResNet50 model was 93.00\%. When the input image width was 1024, the Acc performance of the ResNet50 model was 94.19\%, with no substantial difference between the two. However, when the input image width was 256 or smaller, and the test set was data from after 2015, the model's performance significantly dropped by 2021 to only 81.12\%. In contrast, for images with a width of 1024 and the test set also from after 2015, the model's Acc performance did not decline rapidly, maintaining a minimum of 84.91\%. We believe that such outcomes occur because a smaller image width after conversion leads to partial data loss, affecting accuracy. However, a larger image width significantly increases the model's computational load, resulting in very low efficiency. Therefore, to balance the two, subsequent experiments all opted for an input image width of 1024, avoiding data precision loss while ensuring computational efficiency.

\begin{table}
\vspace{-1.0em}
\tiny
\renewcommand{\arraystretch}{1}
\centering
\caption{From 2015 to 2021, The ResNet50 Model Has Accuracy Results For Image Widths Of 256 and 1024, Respectively.(\%) \label{resnet50_for_256and1024}}
\setlength{\tabcolsep}{4pt}
\begin{tabular}{llllllll}
\hline
\diagbox{Width}{Acc(\%)}{Year} & 2015  & 2016  & 2017  & 2018  & 2019  & 2020   & 2021 \\
\hline
width-256  & 93.00 & 91.76 & 88.27 & 85.80 & 83.87 & 83.43  & 81.12  \\
width-1024 & 94.19 & 91.92 & 84.92 & 86.31 & 85.89 & 87.30  & 88.61  \\
\specialrule{0.05em}{2pt}{2pt}
\end{tabular}
\vspace{-2.0em}
\end{table}

% 在固定了输入的图像宽度为1024之后，为了真正的测试我们模型的有效性，我们将Prompt-Time 2015年的数据作为训练集和测试集，数据足够测试数据驱动的深度学习检测方法中所设计模型的性能可伸缩性和适应性方面。我们对比了其他的深度学习视觉模型, 表ref{malware_detection_result}显示了我们测试的模型和其对应的结果。从表中我们可以观察到我们的，ResNeXt+ \cite{2024resNeXt}是当前恶意软件检测在图像领域最强的方法，我们复现了该方法中表现的最佳的几个模块(e.g., ResNeXt+SE, ResNeXt+CA)。ResNeXt+CA在Prompt-Time 2015 Year数据集上的表现只有。我们的方法PromptSAM+ResNet101达到了最state of the art的效果，达到了95.81%的Accuracy，93.22%的precision以及93.30%的f1-score。除此之外，我们的FPR(误报率)结果为2.95%，我们的FNR(漏报率)结果为6.63%，超越了一众的深度学习模型，达到了最sota的效果。
After fixing the input image width to 1024, to truly test the effectiveness of our model, we used the 2015 data from the Prompt-Time dataset as both the training and testing sets. This data was sufficient to assess the scalability and adaptability of the performance of models designed within data-driven deep learning detection methods. We compared our model with other deep learning visual models, and Table \ref{malware_detection_result} displays the tested models and their corresponding results. From the table, we observed that ResNeXt+ \cite{resNeXt+2024} is currently the strongest method in the domain of malware detection using images; we replicated the best performing modules within this method (e.g. ResNeXt + SE, ResNeXt + CA). The performance of ResNeXt+CA in the Prompt-Time 2015 Year dataset was only 84.30\% and 57.60\% F1-Score. Our method, PromptSAM+ResNet101, achieved state-of-the-art results, with an accuracy of 95. 81\%, precision of 93. 22\%, and an F1 score of 93. 30\%. Furthermore, our False Positive Rate (FPR) was 2.95\%, and our False Negative Rate (FNR) was 6.63\%, surpassing a host of deep learning models to achieve the most state-of-the-art results.

\begin{table}[!t]
\tiny
\renewcommand{\arraystretch}{1}
\setlength{\tabcolsep}{5pt}
\caption{effectiveness of the malware detection methods in dataset prompt-time 2015 year(\%)} 
\label{malware_detection_result}
\centering
\begin{tabular}{cccccccp{1pt}}
\hline
\textbf{Methods} & \textbf{Acc}  & \textbf{Pre}  & \textbf{Rec} & \textbf{F1}  & \textbf{FPR} & \textbf{FNR} \\
\hline
Xception   \cite{xceptionForMalare2019}  &  91.04     &   73.91   &  82.74  & 78.08 & 6.98  & 17.26  \\
ResNet18   \cite{he2016deep} &  94.45    &    85.78   &   88.05   &  86.90  & 3.49  & 11.95          \\ 
ResNet50   \cite{he2016deep} &  94.19    &    84.65   &   85.40   &  85.02  & 3.70  & 14.60          \\ 
Resnet101  \cite{he2016deep} &  94.79    &    84.23   &   89.82   &  86.94  & 4.02  & 10.18          \\
VGG16      \cite{vggForMalware2018} &  94.19  & 84.34 &   85.84   &  85.09  & 3.81  & 14.16          \\
VIT        \cite{vit2020}    &  89.30    &    77.01   &   63.72   &  69.73  & 4.56  & 36.28          \\
ResNeXt101 \cite{resNeXt+2024}    &  87.71    &    73.03   &   57.52   &  64.36  & 5.07  & 42.48     \\
ResNeXt101+CA \cite{resNeXt+2024} &  84.30    &    60.10   &   55.31   &  57.60  & 8.77  & 44.69     \\  
ResNext101+SE \cite{resNeXt+2024} &  87.37    &    67.89   &   65.49   &  66.67  & 7.40  & 34.51     \\  
\specialrule{0.05em}{2pt}{2pt}        
\textbf{PromptSAM+ResNet50}  &  \textbf{95.99} &  \textbf{90.13} & \textbf{88.94} & \textbf{89.53} & \textbf{2.33} & \textbf{11.06 }   \\
\textbf{PromptSAM+ResNet101} &  \textbf{95.81} &  \textbf{91.27} & \textbf{92.29} & \textbf{91.74} & \textbf{2.56} & \textbf{7.71}      \\
\specialrule{0.05em}{2pt}{2pt}
\end{tabular}
\vspace{-0.5cm}
\end{table}

\subsection{Malware Family Classification In Two Platform And Results Analysis}\label{family_classification}
For the metrics used in malware family classification, we employ Accuracy, Precision, Recall, and F1-score, among others. It is particularly noteworthy that for Accuracy in malware family classification, we calculate the value as global Accuracy. This means that for all families, the classifier's proportion of correctly classified samples to the total number of samples across all families is represented. This is reflected in the confusion matrix as the ratio of the sum of all values on the main diagonal to the sum of all values in the matrix, as represented by the formula \ref{global_acc_metric_def}. Here, $C_{i,i}$ represents the number of correctly classified samples, while $C_{i,j}$ denotes the number of malware samples from family i that were classified by the classifier as belonging to family j, where $i,j\in(1,2,... N)$, and N is the number of all malware families in the dataset.

\begin{equation}
\label{global_acc_metric_def}
    Accuracy_{global} = \frac{\Sigma_{i=1}^{N}{C_{i,i}}}{\Sigma_{i=1,j=1}^{N}C_{i,j}}
\end{equation}

\begin{figure}[!t]
\centering
\includegraphics[scale=0.6]{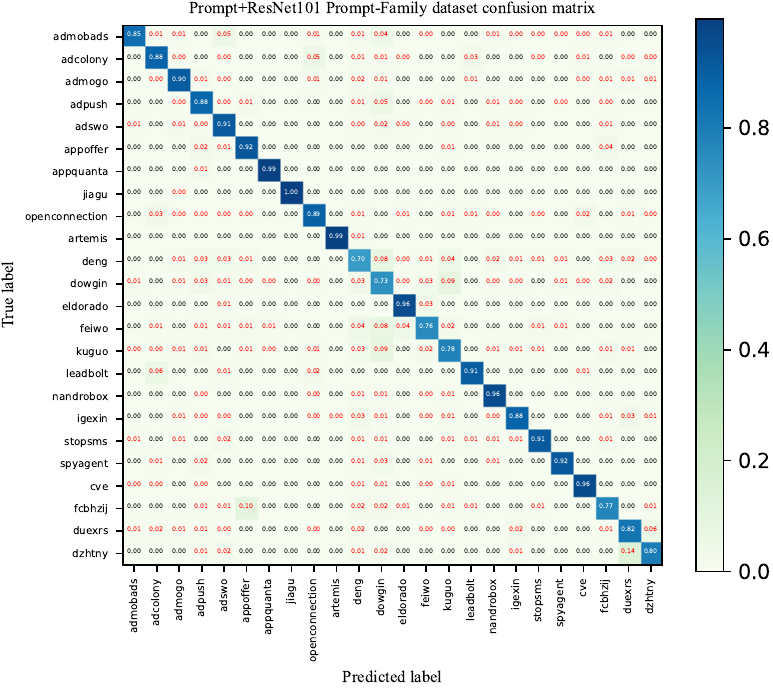}
\caption{The Confusion Matrix of Prompt-Family dataset with PromtSAM+ResNet101.}
\label{fig_Prompt_Family_Confusion_Matrix}
\end{figure}

% \textit{Experiments on Android Platform Dataset:} 我们首先对Android平台恶意软件进行测试，结果如表\ref{our_family_classification_all_family_result_table}所示。从表\ref{our_family_classification_all_family_result_table}可以看出，我们的PromptSAM+ResNet101在Android恶意软件数据集Prompt-Family平均精确率为87.74%，召回率为87.09%，F1-score为87.42%。具体来说，在24个恶意软件家族中，大多数都是正确的，大部分家族分类的准确率接近90%或更高，例如artemis, jiagu, appquanta等家族精确率达到了98%以上。同时, 我们绘制了PromptSAM+ResNet101在Prompt-Family数据集分类结果的混淆矩阵，如图\ref{fig_Prompt_Family_Confusion_Matrix}所示。对于Prompt-Family数据集，对角线值上的值大多显示在0.88左右，离主对角线值显示的值较低。观察表\ref{our_family_classification_all_family_result_table}, 我们发现有三类家族的平均分类精确率低低于80%，它们是‘deng’, ‘dowgin’, ‘feiwo’以及‘kuguo’。通过观察分类结果混淆矩阵\ref{fig_Prompt_Family_Confusion_Matrix}我们发现:(1).恶意软件家族‘deng’ 被错误地分类为‘feiwo’类(占比4%以上)、dowgin(占比3%以上)、kuguo类(占比3%)和igexin(占比3%以上)。(2).恶意软件家族‘dowgin’被错误地分类为‘kuguo’类(占比9%以上)、'dowgin'类(占比8%以上)、‘adpush’类(占比5%以上)和‘admobads’类(占比4%以上)。(3).恶意软件家族‘feiwo’被错误地分类为‘downgin’类（占比3%以上）(4).恶意软件家族‘kuguo’被错误地分类为‘dowgin’类（占比9%以上）、'deng'类（占比4%以上）。实际上，‘deng’, ‘dowgin’, ‘feiwo’以及‘kuguo’这四个恶意软件家族同属一个类型(Type)，它们都属于riskware类型。通过人工分析，我们发现转换后的四个恶意软件家族图像。非常相似。它们很难被人眼识别。Note that some of the malware samples, including 以上的四个恶意软件家族在我们的数据集中，具有相似的结构和图案。因此，基于深度学习模型包括我们的模型也很难将它们区分。我们还使用其他的深度学习模型在Prompt-Family数据上进行实验，实验结果如表\ref{deep_learning_models_for_three_datasets}所示， 在Prompt-Family数据集上，我们在一众深度学习模型中取得了最好的结果。
\textit{Experiments on Android Platform Dataset:} We initially conducted tests on Android platform malware, and the results are shown in Table \ref{our_family_classification_all_family_result_table}. From Table \ref{our_family_classification_all_family_result_table}, it can be seen that our PromptSAM+ResNet101 achieved an average precision of 87.74\%, a recall of 87.09\%, and an F1-score of 87.42\% on the Android malware dataset Prompt-Family. Specifically, among the 24 malware families, most were accurately classified, with most family classification accuracies nearing 90\% or higher, such as $\mathbf{`artemis'}$, $\mathbf{`jiagu'}$, and $\mathbf{`appquanta'}$, where precision reached above 98\%. Additionally, we plotted the confusion matrix for the PromptSAM+ResNet101 on the Prompt-Family dataset, as shown in Figure \ref{fig_Prompt_Family_Confusion_Matrix}. For the Prompt-Family dataset, values on the diagonal mostly appear around 0.88, with lower values off the main diagonal. Observing Table \ref{our_family_classification_all_family_result_table}, we found that three family categories had an average classification precision below 80\%, which are $\mathbf{`deng’}$, $\mathbf{`dowgin’}$, $\mathbf{`feiwo’}$, and $\mathbf{`kuguo’}$. By examining the classification results in the confusion matrix in Fig. \ref{fig_Prompt_Family_Confusion_Matrix} we discovered: (1) The malware family $\mathbf{`deng’}$ was incorrectly classified as $\mathbf{`feiwo’}$ (more than 4\%), $\mathbf{`dowgin'}$ (more than 3\%), $\mathbf{`kuguo'}$ (about 3\%), and $\mathbf{igexin'}$ (more than 3\%). (2) The malware family $\mathbf{`dowgin’}$ was incorrectly classified as $\mathbf{`kuguo’}$ (more than 9\%), $\mathbf{`dowgin'}$ (more than 8\%), $\mathbf{`adpush’}$ (more than 5\%), and $\mathbf{`admobads’}$ (more than 4\%). (3) The malware family $\mathbf{`feiwo’}$ was incorrectly classified as $\mathbf{`downgin’}$ (more than 3\%). (4) The malware family $\mathbf{`kuguo’}$ was incorrectly classified as $\mathbf{`dowgin’}$ (more than 9\%) and $\mathbf{`deng'}$ (more than 4\%). In fact, the four malware families $\mathbf{`deng’, `dowgin’, `feiwo’,}$ and $\mathbf{`kuguo’}$ belong to the same $\mathbf{Type}$ and are categorized as $\mathbf{riskware}$. Upon manual analysis, we discovered that the images of these four malware families, after conversion, are very similar and challenging to distinguish by the human eye. Note that these four malware families in our dataset have similar structures and patterns. Thus, it is difficult for deep learning models, including ours, to differentiate them. We also conducted experiments using other deep learning models on the Prompt-Family data, with results shown in Table \ref{deep_learning_models_for_three_datasets}. On the Prompt-Family dataset, our model achieved the best results among a host of deep learning models.

\begin{table}[!t]
\vspace{-1.0em}
\tiny
\renewcommand{\arraystretch}{1}
\caption{Prompt-family dataset Indicates the classification result of the PromptSAM+ResNet101 malware family} \label{our_family_classification_all_family_result_table}
\centering
\begin{tabular}{cccccccccp{5pt}}
\hline
\textbf{ID} & \textbf{Family}  & \textbf{Pre}  & \textbf{Rec} & \textbf{F1} & \textbf{ID} & \textbf{Family}  & \textbf{Pre}  & \textbf{Rec} & \textbf{F1}  \\
1   & admobads      &    84.87    &    96.64    &   90.28  & 13  & eldorado   &    95.70    &    87.25   &   91.28    \\ 
2   & adcolony      &    88.24    &    90.30    &   89.26  & 14  & feiwo      &    75.51    &    82.22   &   78.72    \\
3   & admogo        &    90.49    &    92.31    &   91.39  & 15  & kuguo      &    78.25    &    80.60   &   79.41    \\
4   & adpush        &    87.55    &    85.05    &   86.28  & 16  & leadbolt   &    90.57    &    87.27   &   88.89    \\
5   & adswo         &    91.39    &    80.59    &   85.65  & 17  & nandrobox  &    95.89    &    93.65   &   94.75    \\
6   & appoffer      &    92.22    &    87.37    &   89.73  & 18  & igexin     &    88.16    &    94.65   &   91.29    \\
7   & appquanta     &    98.81    &    98.22    &   98.52  & 19  & stopsms    &    91.19    &    93.55   &   92.36    \\
8   & jiagu         &    99.67    &    1.0      &   99.83  & 20  & spyagent   &    91.51    &    92.38   &   91.94    \\
9   & openconnection&    88.74    &    90.54    &   89.63  & 21  & cve        &    95.99    &    95.99   &   95.99    \\
10  & artemis       &    99.03    &    99.51    &   99.27  & 22  & fcbhzij    &    77.01    &    63.81   &   69.79    \\
11  & deng          &    69.93    &    75.62    &   72.67  & 23  & duexrs     &    82.49    &    83.79   &   83.14    \\
12  & dowgin        &    72.98    &    60.54    &   66.18  & 24  & dzhtny     &    79.82    &    78.38   &   79.09    \\
\specialrule{0.05em}{2pt}{2pt}
% Average   &     &       &   &    &  Average   &    87.74    &    87.09     &   87.42         \\   
% \specialrule{0.05em}{2pt}{2pt}
\end{tabular}
\vspace{-1em}
\end{table}

\begin{table}[!t]
\vspace{-1.0em}
\tiny
\renewcommand{\arraystretch}{1}
\setlength{\tabcolsep}{3pt}
\caption{DESCRIPTION OF PROMPT-FAMILY DATASET MALWARE FAMILIES\label{family_dataset_description}}
\centering
\begin{tabular}{cccccccc}
\hline
\textbf{ID} & \textbf{Type} & \textbf{Family} & \textbf{Quantity} & {ID} & \textbf{Type} & \textbf{Family} & \textbf{Quantity}\\
\hline
1  & adware     & admobads  &  1500     & 13 & riskware   & eldorado    & 515       \\ 
2  & adware     & adcolony  &  1500     & 14 & riskware   & feiwo       & 678       \\
3  & adware     & admogo    &  1500     & 15 & riskware   & kuguo       & 1500      \\
4  & adware     & adpush    &  1411     & 16 & riskware   & leadbolt    & 554       \\
5  & adware     & adswo     &  1190     & 17 & riskware   & nandrobox   & 1500      \\
6  & adware     & appoffer  &  955      & 18 & smssend    & igexin      & 1500      \\
7  & adware     & appquanta &  886      & 19 & spr        & stopsms     & 778       \\
8  & downloader & jiagu     &  1498     & 20 & spy        & spyagent    & 528       \\
9  & downloader & openconnection & 1494 & 21 & exploit    & cve         & 1500     \\
10 & monitor    & artemis   &  1032     & 22 & troj       & fcbhzij     & 527      \\
11 & riskware   & deng      &  1416     & 23 & smssend++trojan & duexrs & 1272     \\
12 & riskware   & dowgin    &  1500     & 24 & smssend++trojan & dzhtny & 560      \\
\specialrule{0.05em}{2pt}{2pt}
\textbf{Total}  &           &           &    &            &             & 27294        \\
\specialrule{0.05em}{2pt}{2pt}
\end{tabular}
\vspace{-2em}
\end{table}

% \textit{Experiments on Windows Platform Dataset}:除了对Android平台恶意软件图像进行测试, 我们还对Windows平台恶意软件进行测试。Malimg\cite{first_malware_image}是Nataraj 等人在2011 年通过将Windows恶意软件二进制文件读入一个由8位无符号整数组成的矩阵来创建了Malimg数据集。该矩阵可以被视为一个灰度图像，其值范围在 [0, 255] 之间，其中0表示黑色，1表示白色。Malimg数据集使用bin2png脚本从恶意软件文件中提取二进制图像, 形成 3 通道RGB 格式。接着将垂直长图像调整为2种不同的正方形分辨率（224x224 和 300x300 像素)。为了方便实验，在本实验中我们将Malimg数据集只单独分为训练集和测试集。最后，我们有8394个用于训练的恶意软件样本和945个用于测试的样本。
% \textit{Experiments on Windows Platform Dataset}: In addition to testing Android platform malware images, we also conducted tests on Windows platform malware. The Malimg dataset\cite{first_malware_image} was created by Nataraj et al. in 2011 by reading Windows malware binary files into a matrix composed of 8-bit unsigned integers. This matrix can be viewed as a grayscale image, with values ranging from [0, 255], where 0 represents black and 255 represents white. The Malimg dataset uses the bin2png script to extract binary images from malware files, converting them into 3-channel RGB format. Subsequently, vertically elongated images are adjusted to two different square resolutions (224x224 and 300x300 pixels). For the convenience of experimentation, in this study, we divided the Malimg dataset solely into a training set and a test set. Ultimately, we have 8,394 samples for training and 945 samples for testing.

% \textit{Experiments on Windows Platform Dataset}:除了对Android平台恶意软件图像进行测试, 我们还对Windows平台恶意软件进行测试。Malimg\cite{first_malware_image}是Nataraj 等人在2011 年通过将Windows恶意软件二进制文件读入一个由8位无符号整数组成的矩阵来创建了Malimg数据集。该矩阵可以被视为一个灰度图像，其值范围在 [0, 255] 之间，其中0表示黑色，1表示白色。
\textit{Experiments on Windows Platform Dataset}: In addition to testing Android platform malware images, we also conducted tests on Windows platform malware. The Malimg dataset\cite{first_malware_image} was created by Nataraj et al. in 2011 by reading Windows malware binary files into a matrix composed of 8-bit unsigned integers. This matrix can be viewed as a grayscale image, with values ranging from [0, 255], where 0 represents black and 255 represents white. 

% MaleVis\cite{MaleVis}(Malware Evaluation with Vision) 是一个由 25类Windows恶意软件和1类Windows良性软件类别生成的开放式图像数据集，由9100张用于训练的RGB图像和5126张用于测试的RGB图像组成，属于25个恶意软件类和一个良性的软件类。在这个数据集中，每个类包含350个样本用于训练，每个类别包含不同的样本用于测试。为了实验方便，我们将该数据集中的良性软件分离只进行恶意软件家族分类。
MaleVis\cite{MaleVis} (Malware Evaluation with Vision) is an open image dataset generated from 25 categories of Windows malware and one category of Windows benign software, composed of 9,100 RGB images for training and 5,126 RGB images for testing, belonging to 25 malware classes and one benign software class. In this dataset, each class includes 350 samples for training, with each category comprising different samples for testing. For experimental convenience, we have segregated the benign software from this dataset to focus solely on malware family classification.

% 我们将两个Windows的数据集输入到我们提出的PromtSAM+（Mainly PromptSAM+ResNet50 and PromptSAM+ResNet101）模型中，经过60次训练得到实验结果，结果以准确性、精密度、召回率和f1分数来衡量，数据如表\ref{deep_learning_models_for_three_datasets}所示，结果表明我们提出的方法在恶意软件分类中效果良好。其中在Windows数据集上使用我们提出的模型PromptSAM+对于恶意软件分类来说，取得了非常有竞争力的表现，在Malimg数据集表现为99.48%的准确率、98.68的F1-Score。在MaleVis数据集上PromptSAM+Resnet101结果为96.35%的准确率，96.38%的精确率，96.24的召回率，以及96.31%的F1-Score，这个结果比大多数的深度学习模型都要好。
We inputted two Windows data sets into our proposed PromptSAM+ models (mainly PromptSAM+ResNet50 and PromptSAM+ResNet101) and obtained experimental results after more than 60 training iterations. The results were measured in terms of accuracy, precision, recall, and F1-score, as shown in Table \ref{deep_learning_models_for_three_datasets}. The results indicate that our proposed method performs well in malware classification. Specifically, using our PromptSAM+ model on the Windows datasets for malware classification, we achieved very competitive performance, with an accuracy of 99.48\% and an F1-Score of 98.68\% on the Malimg dataset. On the MaleVis dataset, PromptSAM+ResNet101 achieved an accuracy of 96.35\%, a precision of 96.38\%, a recall of 96.24\%, and an F1-Score of 96.31\%, which is better than most deep learning models.

% 我们也绘制了PromptSAM+ResNet101模型在Malimg以及MaleVis数据集上分类结果的混淆矩阵。PromptSAM+ResNet101模型在Malimg数据集混淆矩阵如图\ref{fig_Prompt_Family_Confusion_Matrix}所示。对于我们的测试集，PromptSAM+ResNet101模型产生的混淆矩阵在对角线上的值大多显示为1，离主对角线值显示为低值，这正是我们本次实验所需要的。我们的最终目标是使对角线上的元素都为1。我们可以看到Swizzor.gen!E与Swizzor.gen!I有较高的错误分类，从他们的家族名字就能发现，它们几乎是一样的家族，模式差异较小，所以他们的代码片段几乎是相似的，即使是人工专家也很容易将两者混淆。PromptSAM+ResNet101的在MaleVis数据集上分类结果的混淆矩阵如图\ref{fig_MaleVis_prompt101_confusion_matrix}所示，对于MaleVis数据集，Sality类被错误地分类为Neshta类(占比3%以上)、Expiro类(占比2%以上)和injector类(占比1%)的情况较多。Sality分类结果的准确率在90%以下。Sality类别容易被错误分类的原因是，它集成了Neshta、Expiro和Injector类别的功能\cite{malwareSalityInclude};同时，其感知视场不明显，在为类捕获恶意图像通道感知视场时容易被误分类。
We also created confusion matrices for the PromptSAM+ResNet101 model's classification results on the Malimg and MaleVis datasets. The confusion matrix for the PromptSAM+ResNet101 model on the Malimg dataset is shown in Figure \ref{fig_Prompt_Family_Confusion_Matrix}. For our test set, the confusion matrix generated by the PromptSAM+ResNet101 model mostly shows values of 1 along the diagonal, with low values off the main diagonal, which is precisely what we need for this experiment. Our ultimate goal is to have all elements on the diagonal equal to 1. It is noticeable that `Swizzor.gen!E' and `Swizzor.gen!I' have a higher rate of misclassification. From their family names, it is evident that they are almost the same family with minor pattern differences, leading to similar code fragments and even experts could easily confuse the two. The confusion matrix for PromptSAM+ResNet101 on the MaleVis dataset is shown in Figure \ref{fig_MaleVis_prompt101_confusion_matrix}. For the MaleVis dataset, the `Sality' class was frequently misclassified as the `Neshta' class (more than 3\%), the `Expiro class' (more than 2\%), and the `Injector' class (about 1\%). The accuracy for the `Sality' classification results was below 90\%. The reason for the frequent misclassification of the `Sality' category is that it integrates features from the `Neshta', `Expiro', and `Injector' categories \cite{malwareSalityInclude}. Additionally, its perceptual field is not distinct, making it prone to misclassification when capturing the perceptual field of malicious image channels for the class.

\begin{figure}[!t]
\centering
\includegraphics[scale=0.6]{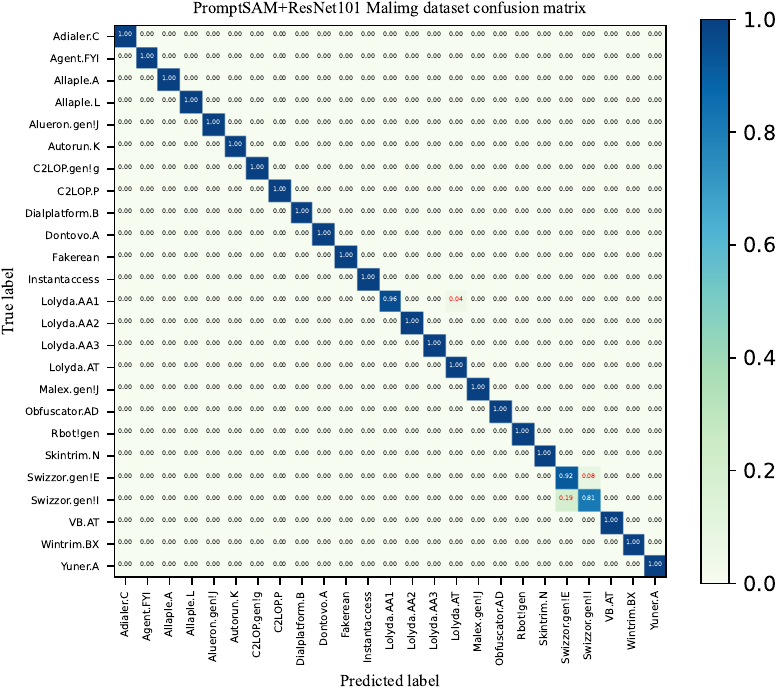}
\caption{The Confusion Matrix of Malimg dataset with Prompt+ResNet101.}
\label{fig_malimg_prompt101_confusion_matirx}
\end{figure}

\begin{figure}[!t]
\vspace{-1.0em}
\centering
\includegraphics[scale=0.6]{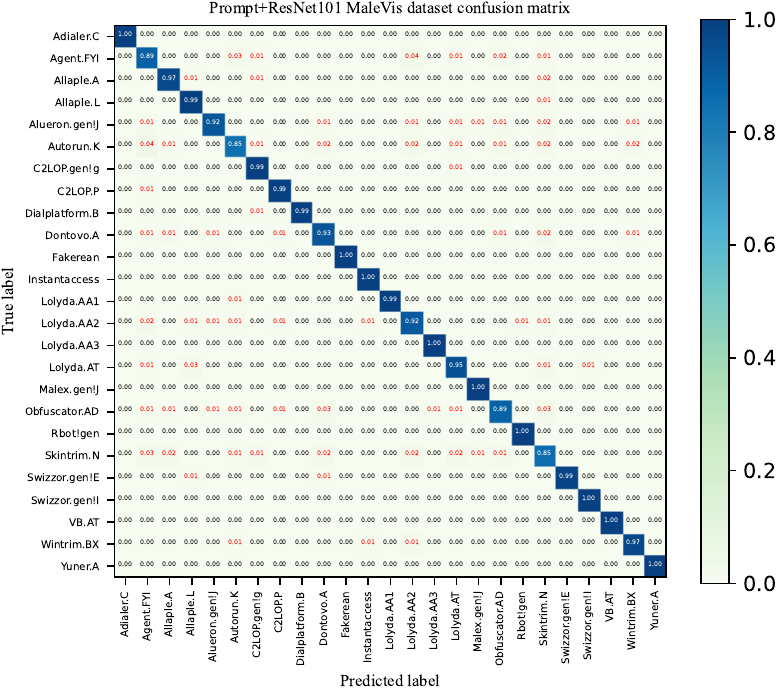}
\caption{The Confusion Matrix of MaleVis dataset with Prompt+ResNet101.}
\label{fig_MaleVis_prompt101_confusion_matrix}
\vspace{-1.0em}
\end{figure}

\begin{table*}
\tiny
\caption{performance comparison of the proposed models with deep learning models for the three datasets(\%) }\label{deep_learning_models_for_three_datasets}
\renewcommand{\arraystretch}{1}
\setlength{\tabcolsep}{5pt}
\centering
\begin{tabular}{l!{\vrule width \lightrulewidth}ccccccccccccc} 
\toprule
\multicolumn{2}{c}{\multirow{2}{*}{\textbf{Method}}}   & \multicolumn{4}{c}{~Malimg Dataset}  & \multicolumn{4}{c}{~ ~MaleVis Dataset}  & \multicolumn{4}{c}{Prompt-Family Dataset}  \\ 
\cmidrule{3-14}
\multicolumn{2}{c}{}                                   & \multicolumn{1}{l}{Accuracy} & \multicolumn{1}{l}{Precision} & \multicolumn{1}{l}{Recall} & \multicolumn{1}{l}{F1-Score} & \multicolumn{1}{l}{Accuracy} & \multicolumn{1}{l}{Precision} & \multicolumn{1}{l}{Recall} & \multicolumn{1}{l}{F1-Score} & \multicolumn{1}{l}{Accuracy} & \multicolumn{1}{l}{Precision} & \multicolumn{1}{l}{Recall} & \multicolumn{1}{l}{F1-Score}  \\ 
\midrule
\multirow{5}{*}{CNN}  & VGG16\cite{vggForMalware2018}                        & 97.44                            & 97.54                            & 97.42                          & 97.48                            & 96.18                            & 94.66                             & 95.67                          & 96.10                            & 85.77                                 & 85.21                                  & 85.09                               & 85.15                                  \\
                      & Inception-v3\cite{inception}                    & 97.65                            & 98.70                             & 98.64                          & 98.67                            & 95.32                            & 95.68                             & 94.99                          & 95.33                            &  83.15                                & 82.38                                  &  81.83                              & 82.10                                \\
                      & ResNet50                       & 97.68                            & 97.61                             & 97.68                          & 97.64                            & 90.36                            & 90.63                             & 89.94                          & 90.28                            & 86.85                                & 86.44                                   & 86.24                               & 86.34                                  \\
                      & Densenet-169\cite{densenet2017}                    & 97.82                            & 97.78                             & 97.83                          & 97.80                            & 95.65                            & 95.86                             & 95.62                          & 95.74                            & 84.86                                 & 83.51                    & 83.76                               & 83.64                                   \\
                      & Xception\cite{xception2017}                        & 96.08                            & 95.76                             & 96.16                          & 95.59                            & 93.00                            & 92.87                             & 91.67                          & 92.27                            & 82.67                           & 82.93                           & 81.51                  & 82.22                                   \\ 
\midrule
VIT                   & Swin Transformer\cite{swinTransformer}                & 93.72                            & 94.13                             & 92.54                          & 93.33                            & 92.24                            & 91.79                             & 90.96                          & 91.37                            & 73.23                                 & 73.48                                 & 72.22                                &72.84                                   \\ 
\midrule
LSTM                  & Deep LSTM\cite{deepLSTM}                       & 96.63                            & 97.32                             & 96.54                          & 96.93                            & 91.31                            & 91.26                             & 90.87                          & 91.06                           & 76.60                               & 75.04                                  & 73.85                                & 74.44                                  \\ 
\midrule
\multirow{2}{*}{Ours} & \textbf{PromptSAM+ResNet50(ours) } & 99.27                         & 98.37                           & 98.37                    & 98.37                            & 95.47                         & 95.45                              & 95.42                      & 95.44                        &  86.74                                & 86.26                         & 86.73                & 86.49      \\
    & \textbf{PromptSAM+ResNet101(ours)}  & \textbf{99.48}                          &\textbf{98.74}                        & \textbf{98.62}                        & \textbf{98.68}            & \textbf{96.35}             & \textbf{96.38}                         & \textbf{96.24}                           & \textbf{96.31}                                  &  \textbf{87.66}                  & \textbf{87.74}                           & \textbf{87.09}                        &  \textbf{87.42}                               \\
\bottomrule
\end{tabular}
\end{table*}

\subsection{Comparison With State-of-the-Art Methods on the malware Family Classification}
% 为了验证我们的模型的性能，我们将我们的恶意软件家族分类实验结果与现有的图像检测恶意软件方法进行了性能比较,我们得到了如表\ref{family_classication_with_sota_table}所示的实验结果。从表\ref{family_classication_with_sota_table}中可以看出，我们的方法在三个数据集的表现结果比大部分方法强。我们的模型PromptSAM+ResNet101在MaleVis数据集的表现相比于ResNeXt+SE\cite{}只差了1.45% F1-Socre, 在Malimg数据集上我们的模型PromptSAM+ResNet101和性能最强的ResNeXt+SE几乎是相同的。我们复现了ResNeXt+(i.e., ResNeXt\cite{}, ResNext+CA, ResNeXt+SE)，但是ResNeXt+CA在Prompt-Family数据集表现只有76.67%的Acc和75.98的F1, ResNeXt+SE在Prompt-Family数据集中表现只有75.23%的Acc和74.34%的F1。相比之下，我们的模型在Prompt-Family获得了87.66%的Acc和87.42% F1, 我们的模型PromptSAM+ResNet101相比于ResNeXt+SE模型在Prompt-Family数据集的结果分别提升了12.43\%\uparrow的Acc和13.08%\uparrow的F1。

% 我们还通过实验将PromptSAM+ResNet50和PromptSAM+ResNet50与现有的深度学习模型例如Swin transformer\cite{swinTransformer}和Deep LSTM\cite{deepLSTM}进行了比较，在三个数据集的实验结果表明我们提出的模型具有较高的精度和泛化能力。
To validate the performance of our model, we compared the results of our malware family classification experiments with existing image-based malware detection methods, and we obtained the experimental results shown in Table \ref{family_classication_with_sota_table}. From Table \ref{family_classication_with_sota_table}, it can be seen that our method performs better than most existing approaches across three datasets. Our model, PromptSAM+ResNet101, on the MaleVis dataset, performed only 1.45\% lower in F1-Score compared to ResNeXt+SE\cite{}. On the Malimg dataset, our PromptSAM+ResNet101's performance was nearly identical to that of the highest-performing ResNeXt+SE. We replicated ResNeXt+ (i.e., ResNeXt\cite{}, ResNext+CA, ResNeXt+SE); however, ResNeXt+CA only achieved 76.67\% Acc and 75.98 F1 on the Prompt-Family dataset, and ResNeXt+SE only managed 75.23\% Acc and 74.34\% F1 on the same dataset. In contrast, our model achieved 87.66\% Acc and 87.42\% F1 on the Prompt-Family, showing our model, PromptSAM+ResNet101, had an increase of 12.43\% $\uparrow$ in Acc and 13.08\% $\uparrow$ in F1 compared to the ResNeXt+SE model on the Prompt-Family dataset.

We also compared PromptSAM+ResNet50 and PromptSAM+ResNet101 with existing deep learning models such as the Swin Transformer\cite{swinTransformer} and Deep LSTM\cite{deepLSTM} through experiments. The results from the three datasets indicate that our proposed models have higher accuracy and generalization capabilities.

\begin{table*}
\tiny
\renewcommand{\arraystretch}{1}
\setlength{\tabcolsep}{5pt}
\caption{performance comparison of the proposed model with other methods in malware family classification on three datasets(\%)}\label{family_classication_with_sota_table}
\centering
\begin{tabular}{ccccccccccccc} 
\toprule
\multirow{2}{*}{\textbf{Method}} & \multicolumn{4}{c}{~Malimg Dataset}                                                                                                      & \multicolumn{4}{c}{~ ~MaleVis Dataset}                                                                                                   & \multicolumn{4}{c}{Prompt-Family Dataset}                                                                                                 \\ 
\cmidrule{2-13}
\ & \multicolumn{1}{l}{Accuracy} & \multicolumn{1}{l}{Precision} & \multicolumn{1}{l}{Recall} & \multicolumn{1}{l}{F1-Score} & \multicolumn{1}{l}{Accuracy} & \multicolumn{1}{l}{Precision} & \multicolumn{1}{l}{Recall} & \multicolumn{1}{l}{F1-Score} & \multicolumn{1}{l}{Accuracy} & \multicolumn{1}{l}{Precision} & \multicolumn{1}{l}{Recall} & \multicolumn{1}{l}{F1-Score}  \\ 
\midrule
Natraaj, GIST+kNN, 2011 \cite{first_malware_image}          & 97.18                            & 96.53                             & 96.85                          & 96.71                            & 91.69                            & 92.36                             & 89.58                          & 90.95                            & -                                & -                                 & -                              & -                                 \\
Narayanan, PCA+kNN, 2016 \cite{PCA+kNN}         & 97.34                            & 96.71                             & 97.04                          & 96.87                            & 91.24                            & 91.18                             & 90.74                          & 90.96                            & -                                & -                                 & -                              & -                                 \\
Vinayakumar, CNN+LSTM, 2019 \cite{CNN+LSTM}      & 96.30                            & 96.30                             & 95.82                          & 96.06                            & 86.29                            & 86.85                             & 86.28                          & 86.56                            & -                                & -                                 & -                              & -                                 \\
Cui, CNN, 2019 \cite{cnnForMalware2018}                   & 94.50                            & 94.64                             & 94.31                          & 94.47                            & 92.13                            & 92.09                             & 91.89                          & 91.99                            & -                                & -                                 & -                              & -                                 \\
Singh, ResNet50, 2019 \cite{resnet50ForMalware2019}            & 96.08                            & 95.76                             & 96.16                          & 95.59                            & 93.00                            & 92.87                             & 91.67                          & 92.27                            & -                                & -                                 & -                              & -                                 \\
Luo, LBP+TF, 2017 \cite{LBP+TF2017}                & 93.72                            & 94.13                             & 92.54                          & 93.33                            & 92.24                            & 91.79                             & 90.96                          & 91.37                            & -                                & -                                 & -                              & -                                 \\
Ma, ACNN, 2019 \cite{ACNN2019}                   & 96.63                            & 97.32                             & 96.54                          & 96.93                            & 91.31                            & 91.26                             & 90.87                          & 91.06                            & -                                & -                                 & -                              & -                                 \\
Gibert, AlexNet, 2016 \cite{alexnet2016}            & 95.33                            & 95.02                             & 95.76                          & 94.89                            & 90.59                            & 91.43                             & 89.79                          & 90.60                            & -                                & -                                 & -                              & -                                 \\
Roseline, DRFP, 2020 \cite{DRFP2020}             & 98.65                            & 98.86                             & 98.63                          & 98.74                            & 97.43                            & 97.53                             & 97.32                          & 97.42                            & -                                & -                                 & -                              & -                                 \\
He, ResNeXt, 2024 \cite{resNeXt+2024}            & 99.14                            & 98.71                            & 98.72                        & 98.70                            & 97.61                            & 97.67                        & 97.54                            & 97.56                            & 73.43                          &73.33                             &72.06                             &72.69                                  \\
He,~ResNeXt+CA, 2024 \cite{resNeXt+2024}            & 99.25                            & 98.93                            & 98.88                          & 98.88                            & 97.34                            & 97.34                          & 97.29                            & 97.29                            &    76.67                          & 76.68                            & 75.30                            &    75.98                         \\
He,~ResNeXt+SE, 2024 \cite{resNeXt+2024}             & 99.57                            & 99.37                             & 99.36                          & 99.36                            & 97.83                            & 97.84                             & 97.81                          & 97.80                            &  75.23                                &  75.20                              & 73.50                             & 74.34                                   \\ 
\midrule
\textbf{PromptSAM+ResNet50(ours) }           & 99.27                            & 98.37                             & 98.37                          & 98.37                            & 95.47                                 & 95.45                                & 95.42                      & 95.44                        &  86.74                                 &   86.26                             & 86.73                   & 86.49       \\
\textbf{PromptSAM+ResNet101(ours)}           & 99.48                            & 98.74                                 & 98.62                               & 98.68                                  & 96.35                                 &96.38                                   & 96.24                               &96.31                                  & \textbf{ 87.66 }                               &    \textbf{87.74 }                          & \textbf{87.09  }                             &  \textbf{87.42 }                              \\
\bottomrule
\end{tabular}
\vspace{-2.5em}
\end{table*}

% ******************************这段不要了************************************
% \subsection{Evaluating Obfuscation Resistance}
% % 我们评估DexRay和Drebin对应用程序混淆的适应能力. Android的混淆有三种：简单策略混淆、复杂策略混淆以及组合混淆,表3给出了简单混淆和非简单混淆以及混合方式混淆的类型和解释。
% \textit{Robustness to Code Obfuscation:} Android app obfuscation is known as the process of modifying an APK so that it is no longer useful to a hacker while remaining fully functional\cite{he2023msdroid}. Android obfuscation comes in three types: trivial strategies, non-trivial strategies, and combined strategies\cite{obfuscations_icse2018}. Table VI provides the types and explanations of simple and non-simple obfuscation.
% ******************************这段不要了*************************************

\subsection{Evaluating PromptSAM+ Slow-Aging Effectiveness}
% 如表\ref{data_drift_regular_and_enhance_table}所示，我们将Prompt-Time中的2015数据用于训练和，2015-2021用于测试，在Regular ResNet101模型检测中，在测试集为2015年的时候模型的性能指标达到最好的状态，但是随时后续测试年份时间的推移，模型的性能产生了极大的衰退，在2020年测试集中Pre以及F1-Score分别下降到只有45.68%，25.96%, 并且误报率，漏报率极大提高。所以如果一个模型的训练时间超过两年，它的使用价值大大降低，在最坏的情况下，没有经过更新完全无法使用，即使模型的Acc指标尚可，但是其漏报和误报严重，极大提高了人力的损耗，失去了其使用价值。为了达到能够对新产生的恶意软件及其家族可以有持续检测目的，就需要完全重新训练一个模型，一方面需要依赖大量重新标记好的数据，另一方面需要花费大量的时间成本和金钱成本去训练一个新的模型，使得成本提高。为此我们提出的模型缓解了部分的问题。
As shown in Table \ref{regular_enhance_table}, we used data from 2015 in Prompt-Time for training, and the period from 2015-2021 for testing. In the Regular ResNet101 model testing, the performance metrics were optimal for the 2015 test set, but as testing proceeded into subsequent years, the model's performance significantly deteriorated. By the 2020 test set, the Precision and F1-Score had plummeted to just 45.68\% and 29.96\%, respectively, and both the false positive rate and false negative rate had greatly increased. Therefore, if a model's training exceeds two years, its utility significantly diminishes, and in the worst case, it becomes completely unusable without updates, despite acceptable accuracy metrics. The serious issues of missed detections and false alarms greatly increase human resource losses, stripping the model of its utility. To achieve ongoing detection capabilities for newly emerging malware and their families, a complete retraining of the model is necessary. This requires reliance on a large amount of newly labeled data and entails substantial time and financial costs, thus increasing the overall cost. Our proposed model alleviates some of these issues.

% 为了便于不同时间衰减图的比较，我们使用了一个新的度量，即\textit{时间下面积}。\textit{AUT}是由Tesseract \cite{pendlebury2019tesseract}提出的度量标准，它定义了每个图中曲线下的面积，以表示模型的可持续性，如式1所示。\eqref{AUT_def}:
To facilitate the comparison of different time decay plots, we use a new metric, \textit{Area Under Time}. \textit{AUT} is a metric proposed by Tesseract \cite{pendlebury2019tesseract}, which defines the area under the curve in each figure to represent the model’s sustainability as shown in Equation \eqref{AUT_def}:
\begin{equation}
\label{AUT_def}
AUT(f,N) = \frac{1}{N-1} \sum_{k=1}^{N-1} \frac{[f(x_{k+1})+f(x_k)]} {2}
\end{equation}

% 其中$mathcal{f}$是性能指标(例如F1分数，精度，召回率等)，N是测试槽的数量，$f(x_k)$是在时间k评估的性能指标，在我们的例子中，最终指标是AUT(metrics,7年)。 %一般情况下，接近1的AUT指标意味着随着时间的推移性能会更好( An AUT metric that is closer to 1 means better performance over time. )。将f设置为Accuracy、Precision、和F1-score、FNR、FPR。
where $f$ is the performance metric (e.g. F1 score, Precision, Recall, etc.),  N is the number of test slots, and $f(x_k)$ is the performance metric evaluated at the time k, and in our case, the final metric is AUT(metrics,7yr). The AUTs of all models with f set to Accuracy, Precision, and F1-score, FNR, FPR.

% 我们在表\ref{regular_enhance_table}中展示了常规的分类器(Regular)和应用PromptSAM增强分类器的结果。这里有两点值得注意。首先，有PromptSAM增强分类器的AUT比没有PromptSAM的分类器的AUT要高。这表明PromptSAM增强之下确实可以减缓六种种不同分类器的模型老化，无论它们是否具有进化意识。其次，使用PromptSAM增强的模型的老化减缓是显著的:例如，在增强了ResNet101之后，对于原始分类器，在相同年份的情况下，只用一个年份数据进行训练的模型可以获得更好的性能，F1平均增加了10.38%, FPR和FNR分别平均减少了0.14%和10.60%。对于regular Vit分类器模型，平均的f1只有43.77%，并且FNR高达67.52%, 使用PromptSAM增强之后, PromtSAM+Vit的f1平均提高17.24%，FNR平均下降了17.84%, 即使Regular Vit的FPR平均值为为2.3%，PromptSAM+Vit对于FPR也能降低0.25%。PromptSAM+VGG16相比于Regular VGG16的FNR也下降了13.08%, 但是Precision也降低了，造成这样结果的原因可能是由于VGG16的模型参数太庞大。我们测试每个分类器的AUT(FNR,7yr), 类似地，我们用应用程序训练分类器2015年至2021年对分类器进行测试，结果如表\ref{AUT_FNR_table}所示。对于ResNet50、ResNet101、Vit、VGG16、Inception-V3和CNN, PromptSAM+可以分别降低8.92%、11.86%、11.05%和1.42%的AUT，并且减少了重复训练的频率, 所以PromptSAM可以节省大量的人力分析。
In Table \ref{regular_enhance_table}, we present the results of conventional classifiers (Regular) and those enhanced by PromptSAM. Two points are particularly noteworthy. First, the Area Under the Time (AUT) for classifiers enhanced by PromptSAM is higher than that for classifiers without PromptSAM. This indicates that PromptSAM enhancement can indeed mitigate the aging of six different classifier models, regardless of whether they are evolution-aware. Second, the aging mitigation achieved using PromptSAM-enhanced models is significant: for example, after enhancing ResNet101, compared to the original classifier, models trained with data from only one year achieved better performance, with an average increase in F1-score of 10.38\%, and average reductions in False Positive Rate (FPR) and False Negative Rate (FNR) of 0.14\% and 10.60\%, respectively. For the regular Vit classifier model, the average F1-score was only 43.77\%, and the FNR was as high as 67.52\%. After enhancement with PromptSAM, the PromptSAM+Vit's average F1-score improved by 17.24\%, and the average FNR decreased by 17.84\%, even though the average FPR for Regular Vit was 2.3\%, PromptSAM+Vit also reduced FPR by 0.25\. PromptSAM+VGG16, compared to Regular VGG16, also showed a decrease in FNR by 13.08\%, but Precision decreased, which might be due to the large model parameters of VGG16. We tested each classifier's AUT (FNR, 7yr), similarly, training classifiers from 2015 to 2021 and testing them, as shown in Table \ref{AUT_FNR_table}. For ResNet50, ResNet101, Vit, VGG16, Inception-V3, and CNN, PromptSAM+ was able to reduce the AUT by 8.92\%, 11.86\%, 11.05\%, and 1.42\% respectively, and also reduced the frequency of retraining, thus PromptSAM can save a significant amount of manpower in analysis.

\begin{table}
\label{AUT_FNR_table}
\centering
\tiny
\renewcommand{\arraystretch}{0.7}
\setlength{\tabcolsep}{3pt}
\caption{AUT(FNR,7yr) of original (w/o) and enhanced(w/) classifiers.(\%)}
\begin{threeparttable}          %这行要添加
\begin{tabular}{ccccccccccccccccc} 
\toprule
\multicolumn{2}{c}{\textbf{ResNet50 }} &  & \multicolumn{2}{c}{\textbf{ResNet101 }} &  & \multicolumn{2}{c}{\textbf{Vit}} &  & \multicolumn{2}{c}{\textbf{\textbf{VGG16}}} &  & \multicolumn{2}{c}{\textbf{\textbf{CNN}}} &  & \multicolumn{2}{c}{\textbf{\textbf{Inception-v3}}}  \\ 
\cline{1-2}\cline{4-5}\cline{7-8}\cline{10-11}\cline{13-14}\cline{16-17}
w/o   & w/                             &  & w/o   & w/~                             &  & w/o   & w/                       &  & w/o   & w/                                  &  & w/o   & w/                                &  & w/o   & w/                                          \\ 
\cline{1-2}\cline{4-5}\cline{7-8}\cline{10-11}\cline{13-14}\cline{16-17}
58.33 & 49.41                          &  & 57.01 & 45.15                           &  & 69.18 & 52.48                    &  & 52.26 & 41.21                               &  & 56.53 & 48.46                             &  & 54.88 & 53.46                                       \\
\bottomrule
\end{tabular}
 \begin{tablenotes}
        \tiny
        \item[1] w/o denotes the classifier without PromptSAM+, i.e. the original classifier. 
        \item[2] w/ denotes the classifier enhanced with PromptSAM+.
\end{tablenotes}
\end{threeparttable}
\vspace{-2.0em}
\end{table}
We tested each classifier's AUT (FNR, 7yr), similarly, training classifiers from 2015 to 2021 and testing them, as shown in Table \ref{AUT_FNR_table}. For ResNet50, ResNet101, Vit, VGG16, Inception-V3, and CNN, PromptSAM+ was able to reduce the AUT by 8.92\%, 11.86\%, 11.05\%, and 1.42\% respectively, and also reduced the frequency of retraining, thus PromptSAM can save a significant amount of manpower in analysis.

% \begin{figure}[!t]
% \centering
% \includegraphics[scale=0.65]{fig_resnet101_decrease.pdf}
% \caption{ResNet101 .}
% \label{fig_resnet101_derease}
% \end{figure}

\begin{table*}
\label{regular_enhance_table}
\tiny
\centering
\renewcommand{\arraystretch}{1.3}
\setlength{\tabcolsep}{4pt}
\caption{Comparisons of the regular and enhanced models (\%)}
\begin{tabular}{c|cccc|cccc|cccc|cccc|cccc|cccc} 
\toprule
\multirow{2}{*}{\begin{tabular}[c]{@{}c@{}}Testing\\ Years\end{tabular}} & \multicolumn{4}{c|}{Regular ResNet50}                                                                         & \multicolumn{4}{c|}{PromptSAM+ResNet50}                                                                                                         & \multicolumn{4}{c|}{Regular ResNet101}                                                                        & \multicolumn{4}{l|}{PromptSAM+ResNet101}                                                                                                             & \multicolumn{4}{c|}{Regular Inception-V3}                                                                     & \multicolumn{4}{l}{PromptSAM+Inception-V3}                                                                                                      \\ 
\cline{2-25}
                                                                         & Pre                       & F1                        & FPR                      & FNR                        & Pre                             & F1                               & FPR                                 & FNR                                  & Pre                       & F1                        & FPR                      & FNR                        & Pre                                  & F1                               & FPR                                 & FNR                                  & Pre                       & F1                        & FPR                      & FNR                        & Pre                              & F1                              & FPR                                 & FNR                                  \\ 
\hline
2015                                                                     & 87.22                     & 87.42                     & 3.07                     & \multicolumn{1}{c}{12.39}  & 90.13                           & 89.53                            & 2.33                                & 11.06                                & 84.23                     & 86.94                     & 4.02                     & \multicolumn{1}{c}{10.18}  & 91.27                                & 91.74                            & 2.56                                & 7.71                                 & 82.74                     & 82.74                     & 4.12                     & \multicolumn{1}{c}{17.26}  & 87.01                            & 87.01                           & 3.7                                 & 11.06                                \\
2016                                                                     & 87.76                     & 77.65                     & 2.45                     & 30.36                      & 89.66                           & 80.89                            & 2.15                                & 26.32                                & 84.08                     & 75.45                     & 3.27                     & 31.58                      & 84.58                                & 81.01                            & 3.58                                & 22.27                                & 87.70                     & 75.58                     & 2.35                     & 33.60                      & 78.46                            & 78.46                           & 2.15                                & 29.96                                \\
2017                                                                     & 92.59                     & 45.45                     & 0.63                     & \multicolumn{1}{c}{69.88}  & 94.08                           & 71.32                            & 0.95                                & 42.57                                & 93.02                     & 47.76                     & 0.63                     & \multicolumn{1}{c}{67.87}  & 89.80                                & 66.67                            & 1.59                                & 46.99                                & 98.54                     & 69.95                     & 0.21                     & \multicolumn{1}{c}{45.78}  & 52.99                            & 52.99                           & 0.95                                & 62.65                                \\
2018                                                                     & 89.63                     & 53.19                     & 1.11                     & 62.19                      & 89.10                           & 58.40                            & 1.35                                & 56.56                                & 81.82                     & 55.67                     & 2.38                     & 57.81                      & 83.96                                & 61.93                            & 2.39                                & 50.94                                & 86.49                     & 54.70                     & 1.59                     & 60.00                      & 60.38                            & 60.38                           & 1.03                                & 55                                   \\
2019                                                                     & 91.78                     & 44.22                     & 0.62                     & \multicolumn{1}{c}{70.87}  & 88.24                           & 47.62                            & 1.04                                & 67.39                                & 69.66                     & 38.87                     & 2.81                     & \multicolumn{1}{c}{73.04}  & 73.68                                & 54.29                            & 3.65                                & 57.02                                & 72.63                     & 42.46                     & 2.71                     & \multicolumn{1}{c}{70.00}  & 53.21                            & 53.21                           & 1.04                                & 62.17                                \\
2020                                                                     & 80.95                     & 32.69                     & 0.80                     & 79.52                      & 81.67                           & 43.36                            & 1.10                                & 70.48                                & 45.68                     & 29.96                     & 4.40                     & 77.71                      & 67.39                                & 48.06                            & 3.01                                & 62.65                                & 39.00                     & 29.32                     & 6.11                     & 76.51                      & 37.17                            & 37.17                           & 1.8                                 & 74.7                                 \\
2021                                                                     & 81.52                     & 51.90                     & 1.74                     & \multicolumn{1}{c}{61.93}  & 91.67                           & 60.07                            & 0.82                                & 55.33                                & 75.45                     & 54.07                     & 2.76                     & \multicolumn{1}{c}{57.87}  & 78.26                                & 57.69                            & 2.55                                & 54.31                                & 43.90                     & 39.89                     & 9.40                     & \multicolumn{1}{c}{63.45}  & 55.67                            & 55.67                           & 1.33                                & 58.88                                \\ 
\hline
average                                                                  & \multicolumn{1}{l}{87.35} & \multicolumn{1}{l}{56.08} & 1.49                     & 55.31                      & \multicolumn{1}{l}{89.22}       & \multicolumn{1}{l}{64.46}        & 1.39                                & 47.10                                & \multicolumn{1}{l}{76.28} & \multicolumn{1}{l}{55.53} & 2.90                     & 53.72                      & \multicolumn{1}{l}{81.28}            & \multicolumn{1}{l}{65.91}        & \multicolumn{1}{l}{2.76}            & \multicolumn{1}{l|}{43.13}           & \multicolumn{1}{l}{73.00} & \multicolumn{1}{l}{56.38} & \multicolumn{1}{l}{3.78} & \multicolumn{1}{l|}{52.37} & \multicolumn{1}{l}{60.70}        & \multicolumn{1}{l}{60.70}       & \multicolumn{1}{l}{1.71}            & \multicolumn{1}{l}{50.63}            \\ 
\hline
improve                                                                  & -                         & -                         & -                        & -                          & \textcolor{red}{$\uparrow$1.87} & \textcolor{red}{$\uparrow$8.38}  & \textcolor{green}{$\downarrow$0.10} & \textcolor{green}{$\downarrow$8.20}  & -                         & -                         & -                        & \textbf{-}                 & \textcolor{red}{$\uparrow$5.00}      & \textcolor{red}{$\uparrow$10.38} & \textcolor{green}{$\downarrow$0.14} & \textcolor{green}{$\downarrow$10.6}  & -                         & -                         & -                        & -                          & \textcolor{red}{$\uparrow$13.16} & \textcolor{red}{$\uparrow$4.32} & \textcolor{green}{$\downarrow$2.07} & \textcolor{green}{$\downarrow$1.74}  \\ 
\hline
\multirow{2}{*}{\begin{tabular}[c]{@{}c@{}}Testing\\ Years\end{tabular}} & \multicolumn{4}{c|}{Regular Vit}                                                                              & \multicolumn{4}{c|}{PromptSAM+Vit}                                                                                                              & \multicolumn{4}{c|}{Regular VGG16}                                                                            & \multicolumn{4}{c|}{PromptSAM+VGG16}                                                                                                                 & \multicolumn{4}{c|}{Regular CNN}                                                                              & \multicolumn{4}{c}{PromtptSAM+CNN}                                                                                                              \\ 
\cline{2-25}
                                                                         & Pre                       & F1                        & FPR                      & FNR                        & Pre                             & F1                               & FPR                                 & FNR                                  & Pre                       & F1                        & FPR                      & FNR                        & Pre                                  & F1                               & FPR                                 & FNR                                  & Pre                       & F1                        & FPR                      & FNR                        & Pre                              & F1                              & FPR                                 & FNR                                  \\ 
\hline
2015                                                                     & \multicolumn{1}{l}{74.33} & \multicolumn{1}{l}{67.31} & \multicolumn{1}{l}{5.10} & \multicolumn{1}{l}{38.50}  & 80.40                           & 84.45                            & 5.18                                & 11.06                                & \multicolumn{1}{l}{84.35} & \multicolumn{1}{l}{85.09} & \multicolumn{1}{l}{3.81} & \multicolumn{1}{l}{14.16}  & 75.74                                & 82.73                            & 6.98                                & 8.85                                 & \multicolumn{1}{l}{83.72} & \multicolumn{1}{l}{81.63} & \multicolumn{1}{l}{3.70} & \multicolumn{1}{l}{20.35}  & 84.32                            & 86.15                           & 3.91                                & 11.95                                \\
2016                                                                     & \multicolumn{1}{l}{79.61} & \multicolumn{1}{l}{60.65} & \multicolumn{1}{l}{3.17} & \multicolumn{1}{l|}{51.01} & 85.58                           & 78.24                            & 3.07                                & 27.94                                & \multicolumn{1}{l}{88.66} & \multicolumn{1}{l}{78.00} & \multicolumn{1}{l}{2.25} & \multicolumn{1}{l|}{30.36} & 86.88                                & 82.05                            & 2.97                                & 22.27                                & \multicolumn{1}{l}{88.17} & \multicolumn{1}{l}{75.75} & \multicolumn{1}{l}{2.25} & \multicolumn{1}{l|}{33.60} & 83.33                            & 78.89                           & 3.79                                & 25.10                                \\
2017                                                                     & \multicolumn{1}{l}{76.70} & \multicolumn{1}{l}{44.89} & \multicolumn{1}{l}{2.55} & \multicolumn{1}{l}{68.27}  & 86.61                           & 58.51                            & 1.8                                 & 55.82                                & \multicolumn{1}{l}{95.00} & \multicolumn{1}{l}{68.38} & \multicolumn{1}{l}{0.74} & \multicolumn{1}{l}{46.59}  & 88.42                                & 76.54                            & 2.33                                & 32.53                                & \multicolumn{1}{l}{91.47} & \multicolumn{1}{l}{62.43} & \multicolumn{1}{l}{1.16} & \multicolumn{1}{l}{52.61}  & 86.13                            & 61.14                           & 2.01                                & 52.61                                \\
2018                                                                     & \multicolumn{1}{l}{71.00} & \multicolumn{1}{l}{33.81} & 2.31                     & 77.81                      & 89.19                           & 56.41                            & 1.27                                & 58.75                                & \multicolumn{1}{l}{94.26} & \multicolumn{1}{l}{52.04} & 0.56                     & 64.06                      & 83.42                                & 63.97                            & 2.62                                & 48.13                                & \multicolumn{1}{l}{77.95} & \multicolumn{1}{l}{44.30} & \multicolumn{1}{l}{2.23} & \multicolumn{1}{l|}{69.06} & 85.33                            & 62.43                           & 2.15                                & 50.78                                \\
2019                                                                     & \multicolumn{1}{l}{82.54} & \multicolumn{1}{l}{35.62} & 1.15                     & \multicolumn{1}{c}{77.29}  & 89.53                           & 48.73                            & 0.94                                & 66.52                                & \multicolumn{1}{l}{86.11} & \multicolumn{1}{l}{41.06} & 1.04                     & \multicolumn{1}{c}{73.04}  & 64.85                                & 54.18                            & 6.04                                & 53.48                                & \multicolumn{1}{l}{72.90} & \multicolumn{1}{l}{46.29} & \multicolumn{1}{l}{3.02} & \multicolumn{1}{l}{66.09}  & 81.31                            & 51.79                           & 2.09                                & 62.01                                \\
2020                                                                     & \multicolumn{1}{l}{71.79} & \multicolumn{1}{l}{27.32} & 1.11                     & 83.13                      & 78.95                           & 40.36                            & 1.2                                 & 72.89                                & \multicolumn{1}{l}{82.22} & \multicolumn{1}{l}{35.07} & 0.80                     & 77.71                      & 54.55                                & 43.48                            & 5.01                                & 63.86                                & \multicolumn{1}{l}{53.33} & \multicolumn{1}{l}{33.20} & \multicolumn{1}{l}{3.50} & \multicolumn{1}{l|}{75.90} & 77.14                            & 45.76                           & 1.60                                & 67.47                                \\
2021                                                                     & \multicolumn{1}{l}{86.79} & \multicolumn{1}{l}{36.80} & 0.72                     & \multicolumn{1}{c}{76.65}  & 90.82                           & 60.34                            & 0.92                                & 54.82                                & \multicolumn{1}{l}{94.05} & \multicolumn{1}{l}{56.23} & 0.51                     & \multicolumn{1}{c}{59.90}  & 71.52                                & 62.07                            & 4.39                                & 45.18                                & \multicolumn{1}{l}{60.00} & \multicolumn{1}{l}{45.43} & \multicolumn{1}{l}{4.90} & \multicolumn{1}{l}{63.45}  & 80.53                            & 58.71                           & 2.25                                & 53.81                                \\
average                                                                  & \multicolumn{1}{l}{77.54} & \multicolumn{1}{l}{43.77} & 2.3                      & 67.52                      & 85.87                           & 61.01                            & 2.05                                & 49.69                                & \multicolumn{1}{l}{89.24} & \multicolumn{1}{l}{59.41} & 1.39                     & 52.26                      & 75.05                                & 66.43                            & 4.33                                & 39.18                                & \multicolumn{1}{l}{75.36} & \multicolumn{1}{l}{55.57} & \multicolumn{1}{l}{2.97} & \multicolumn{1}{l|}{54.44} & 82.59                            & 63.55                           & 2.54                                & 46.25                                \\ 
\hline
improve                                                                  & -                         & -                         & -                        & -                          & \textcolor{red}{$\uparrow$8.33} & \textcolor{red}{$\uparrow$17.24} & \textcolor{green}{$\downarrow$0.25} & \textcolor{green}{$\downarrow$17.84} & -                         & -                         & \textbf{-}               & \textbf{-}                 & \textcolor{green}{$\downarrow$14.18} & \textcolor{red}{$\uparrow$7.02}  & \textcolor{red}{$\uparrow$2.95}     & \textcolor{green}{$\downarrow$13.08} & -                         & -                         & -                        & -                          & \textcolor{red}{$\uparrow$7.22}  & \textcolor{red}{$\uparrow$7.98} & \textcolor{green}{$\downarrow$0.42} & \textcolor{green}{$\downarrow$8.19}  \\
\bottomrule
\end{tabular}
\vspace{-2.0em}
\end{table*}

\subsection{Malware Family Classification Explanation Analysis}
\begin{figure}[!t]
\centering
\includegraphics[scale=0.24]{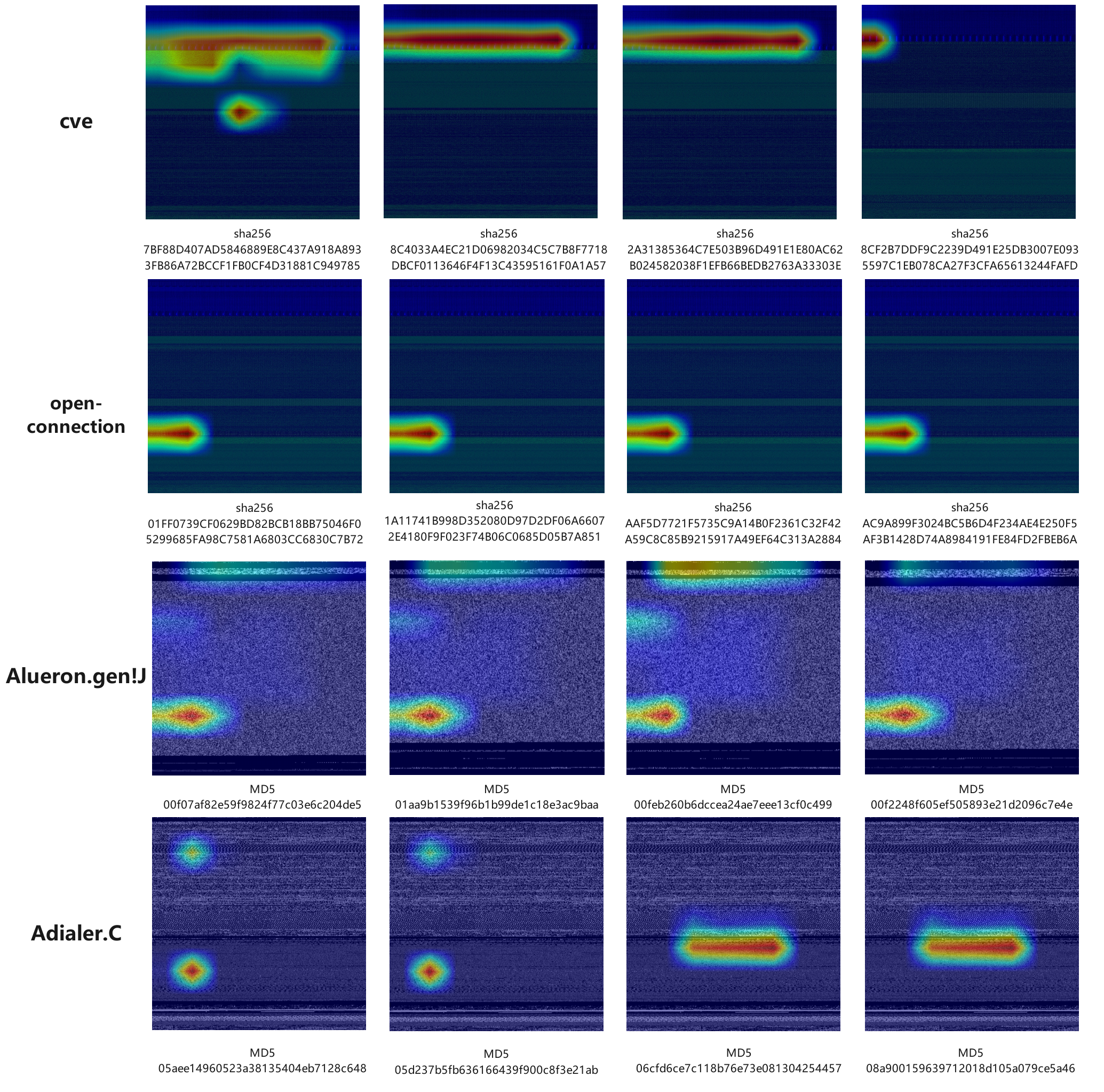}
\caption{PromptSAM+ResNet101 Model Attention Patterns Across 4 Malware Types (each with 4 images).}
\label{fig_family_heat_map}
\vspace{-2.0em}
\end{figure}

In the work presented in this paper, our approach is based on the fact that malware from the same family exhibit similar characteristics, allowing the identification of common malicious components within samples from the same family \cite{gefdroid}. It is understandable that such results would occur; malware authors often reuse code when creating new variants within the same family, resulting in a significant amount of shared code among family members. Consequently, malware from the same family exhibits a high degree of similarity in their compiled bytecode, and the parts of the bytecode that are identical display similar visual texture characteristics when converted into images. Families such as 'cve' and 'openconnection' represent visualizations of Android malware, while 'Alueron.gen!J' and 'Adialer.C' are visualizations of PE malware. Each image is accompanied by the sha256 or MD5 values of the original Android or PE files below it. We utilized the popular Grad-Cam\cite{gradcam} technique with our PromptSAM+ResNet101 model to highlight regions of interest for four types of malware and benign software, with the heatmap shown in Figure \ref{fig_family_heat_map}. We observed that the attention maps are concentrated on the data sections of the bytecode (parts typically storing malicious payloads), and the model's attention for most of the same family is consistently focused on the same locations. Specific byte sequences identified in the attention maps are closely related to the malicious code payloads. Visual analysis can significantly reduce the time and effort required for security analysts to manually investigate suspicious areas within the bytecode.

% \subsection{Few-shot Learning Result}
% 在训练的过程中我们注意到, 以ResNet为代表的预训练残差网络和CNN等计算机视觉网络，在训练的初期模型准确度不高，在只有很少量数据时accuracy等评价指标很差，仅仅只有30%左右，而我们方法在训练少量的样本时，准确度高达86%以上。然而在恶意软件进化过程中，一次性收集大量恶意软件是一件困难的任务，因此我们对本文的模型进行了few-shot实验，以证明在训练集不足的情况下，我们的模型保持了良好的判别能力。我们在Prompt-Family Dataset上进行了实验，设置训练集数量N=50，包含40个正样本和10个负样本。实验结果见表\ref{}。

\subsection{Ablation Study}
%在什么数据集上做了消融实验，消融的是什么

% 本文的模型能够生效主要是提取到带有语义的视觉模型的语义信息，因此我们针对此设计了消融实验。我们在Prompt Time 2015 year作为数据集进行了实验。我们的promptSAM+模型是一个可插入模块，因此我们对不带有语义信息的视觉模型以同样的方式设计了可插入结构。我们采用的是预训练的vit模型和diffusion模型。对于diffusion模型，我们采用的是从UNet中抽取feature进行聚合的方式，与promptSAM模型的结构相似，我们从不同的block中把feature抽取出来，然后上采样到相同维度，再聚合到一起。为了保证与promptSAM模型的一致性，我们也在其后接入ResNet101作为分类头。

% 由于我们的PromptSAM+模型的feature是从SAM的image encoder中提取的，而SAM的image encoder主要由ViT组成。为了确定ViT架构的预训练模型对实验结果的影响，我们设计了对ViT预训练模型提取feature后用ResNet101分类的实验，提取feature的模块与前文中提取SAM的feature模块设计思路完全相同。

% 结果如表\ref{ablation_study_table}所示。对于vit预训练模型对ResNet101分类头，该实验结果表明，vit预训练模型对于实验结果的提升有限，Vit backbone+ResNet101只有33.32%的f1，AUT(FNR,7yr)为79.53%。即promptSAM模型的效果并非来源于vit结构。我们对于不带语义的diffusion预训练模型的实验中，该实验结果表明Diffusion Backbone+ResNet101只有51.71%的f1，并且FNR平均高达47.71%。所以不含语义的复杂视觉模型对恶意软件分类的提升较小。我们的消融实验能够证明，promptSAM模型对已有模型的提升能力来自于其丰富的语义信息。
The effectiveness of our model primarily derives from the semantic information extracted from the visual models with semantics, for which we designed an ablation study. We conducted experiments using the Prompt Time 2015 year dataset. Our PromptSAM+ model functions as an insertable module, so we designed a similar insertable structure for visual models without semantic information, utilizing pretrained ViT and diffusion models. For the diffusion model, we adopted a method of extracting features from UNet and aggregating them, similar to the structure of the PromptSAM model, where features are extracted from different blocks, then upsampled to the same dimension, and aggregated together. To ensure consistency with the PromptSAM model, we also connected a ResNet101 as the classification head.

Since the features of our PromptSAM+ model are extracted from SAM’s image encoder, primarily composed of ViT, we designed experiments to assess the impact of the ViT architecture’s pretrained model on experimental results by extracting features using ResNet101 for classification, with the feature extraction module designed similarly to the module that extracts features from SAM.

The results are shown in Table \ref{ablation_study_table}. For the ViT pretrained model paired with the ResNet101 classification head, the results indicate that the ViT pretrained model contributes limited enhancement to the experimental outcomes; Vit backbone + ResNet101 achieved only a 33.32\% F1-score, and AUT (FNR, 7yr) was 79.53\%. This implies that the efficacy of the PromptSAM model does not originate from the ViT structure. In experiments with the semantically unladen diffusion pretrained model, the results showed that Diffusion Backbone + ResNet101 achieved only a 51.71\% F1-score, with an average FNR as high as 47.71\%. Therefore, complex visual models without semantic content contribute minimally to enhancements in malware classification. Our ablation study demonstrates that the improvement capability of the PromptSAM model stems from its rich semantic information.
\begin{table}
\tiny
\renewcommand{\arraystretch}{1.3}
\setlength{\tabcolsep}{3pt}
\caption{COMPARISON OF DIFFERENT backbone WITH ResNet101 ON Prompt-Time DATASETS FOR MALWARE detection.(\%)}
\label{ablation_study_table}
\begin{tabular}{l|llllll|llllll} 
\hline
                  & \multicolumn{6}{c|}{Diffusion Backbone}       & \multicolumn{6}{c}{Vit Backbone}              \\ 
\hline
metrics      & Acc   & Pre   & Rec   & F1    & FPR   & FNR   & Acc   & Pre   & Rec   & F1    & FPR  & FNR    \\
Average      & 82.99 & 54.76 & 52.29 & 52.71 & 10.12 & 47.71 & 84.81 & 81.79 & 22.59 & 33.32 & 0.97 & 77.41  \\
AUT(m,7yr)   & 82.92 & 55.27 & 50.36 & 51.98 & 9.71  & 49.64 & 84.41 & 82.09 & 20.47 & 31.09 & 0.88 & 79.53  \\
\hline
\end{tabular}
\vspace{-2.0em}
\end{table}

\section{CONLUSION AND FUTURE WORK}\label{conclusion_and_future}
% 在本文中，我们提出了一个基于Prompt Segment Anything Model用于恶意软件检测和恶意软件家族分类的PromptSAM+网络。详细描述了基于Prompt SAM不同的分类头模块的构建和模型实现的细节。PromptSAM+ 首次尝试将基于将带有语义的视觉网络与恶意软件检测任务相结合。通过这种设计，我们可以从大视觉网络SAM中提取到带有语义的视觉模型的语义信息，利用这些语义信息，可以普遍增强现有基图像的恶意软件的分类器的检测能力。我们广泛的评估结果表明，PromptSAM+在性能和实用性方面优于最先进的方法，并且可以缓解恶意软件模型的老化，极大提高模型的可持续使用能力。未来，我们计划通过设计轻量级、高效的模型取代PromptSAM+集成到智能移动设备中，以取代现有的模块。我们还计划研究更高级的机制，以支持模型在面对模型衰减时的进化。
In this paper, we introduce the PromptSAM+ network, a system based on the Prompt Segment Anything Model designed for malware detection and malware family classification. We provide detailed descriptions of the construction and implementation of various classifier head modules based on Prompt SAM. PromptSAM+ represents the first attempt to integrate a semantically enriched visual network with malware detection tasks. Through this design, we extract semantic information from the large visual network SAM, and this semantic information can broadly enhance the detection capabilities of existing image-based malware classifiers. Our extensive evaluations show that PromptSAM+ outperforms state-of-the-art methods in terms of performance and practicality, and can alleviate the aging of malware models, significantly enhancing the model's sustainability. In the future, we plan to develop lightweight, efficient models to replace the current modules integrated into smart mobile devices with PromptSAM+. We also aim to explore more advanced mechanisms to support the model’s evolution in the face of model decay.

\bibliographystyle{IEEEtran}
\bibliography{document.bib}%第二个参数参数是你的bib文件的名字

\vfill
\end{document}